# A machine-learned kinetic energy model for light weight metals and compounds of group III-V elements

Johann Lüder[1,2,3], Manabu Ihara[4], Sergei Manzhos[4,a]

[1] Department of Materials and Optoelectronic Science, National Sun Yat-sen University, 80424, No. 70, Lien-Hai Road, Kaohsiung, Taiwan

[2] Center of Crystal Research, National Sun Yat-sen University, 80424, No. 70, Lien-Hai Road, Kaohsiung, Taiwan

[3] Center for Theoretical and Computational Physics, National Sun Yat-Sen University, Kaohsiung 80424, Taiwan

[4] School of Materials and Chemical Technology, Tokyo Institute of Technology, Ookayama 2-12-1, Meguro-ku, Tokyo 152-8552 Japan

**Abstract**

We present a machine-learned (ML) model of kinetic energy for orbital-free density functional theory (OF-DFT) suitable for bulk light weight metals and compounds made of group III-V elements. The functional is machine-learned with Gaussian process regression (GPR) from data computed with Kohn-Sham DFT with plane wave bases and local pseudopotentials. The dataset includes multiple phases of unary, binary, and ternary compounds containing Li, Al, Mg, Si, As, Ga, Sb, Na, Sn, P, and In. A total of 433 materials were used for training, and 18 strained structures were used for each material. Averaged (over the unit cell) kinetic energy density is fitted as a function of averaged terms of the 4$^{th}$ order gradient expansion and the product of the density and effective potential. The kinetic energy predicted by the model allows reproducing energy-volume curves around equilibrium geometry with good accuracy. We show that the GPR model beats linear and polynomial regressions. We also find that unary compounds sample a wider region of the descriptor space than binary and ternary compounds, and it is therefore important to include them in the training set; a GPR model trained on a small number of unary compounds is able to extrapolate relatively well to binary and ternary compounds but not vice versa.

---

[a] Author to whom correspondence should be addressed. Email: Manzhos.s.aa@m.titech.ac.jp

**Keywords**: orbital-free density functional theory, kinetic energy functional, machine learning, Gaussian process regression

# 1 Introduction

Orbital-free DFT (OF-DFT)[1–4] holds the promise of enabling routine large-scale ($10^6$ atoms and beyond) ab initio materials modeling, as it can achieve near-linear scaling with small prefactors (much smaller than those of order-N DFT,[5–8] for example). This is important on one hand to bring the scale of ab initio simulations closer to reality, in particular for intrinsically large-scale problems such as microstructure-driven properties, doping, interfaces, multicomponent alloys etc., and on the other hand for more efficient utilization of costly CPU resources currently spent on near-cubically scaling Kohn-Sham DFT[9,10] simulations widely performed in applications. Especially with recent advances in enabling excited state calculations with OF-DFT[11–13] and in obtaining mechanistic insight directly from the electron density (as orbitals are not computed in OF-DFT),[14–16] the need to develop accurate enough OF-DFT solutions for use in mainstream applications is obvious. However, OF-DFT is still not applications-ready (beyond some light metals[17–24]) due to the lack of sufficiently accurate kinetic energy functionals (KEF) and partly due to the lack of OF-DFT- suited (local) pseudopotentials (LPP).[21,25–29] This situation, which persisted for decades, is rapidly changing,[4,11,30,30–40] in particular, due to growing and increasingly successful effort in using machine learning (ML) for both the KEF[41–54] and the LPP.[20,55,56]

The reader is referred to the literature,[1–4] in particular to the recent comprehensive review of Ref. [4], for the description of OF-DFT, its advantages and disadvantages vs. KS DFT, and its key equations which will not be repeated here. The reader is also referred to the literature for the description of key machine learning methods, namely, neural networks and kernel methods such as Gaussian process regression and kernel ridge regression,[57–60] which will not be repeated here. One typically machine-learns the Kohn-Sham kinetic energy (KE)

$$E_{kin}^{KS} = -\frac{1}{2}\sum_{i=1}^{N_{el}} \int \psi_i^*(\boldsymbol{r})\Delta\psi_i(\boldsymbol{r})d\boldsymbol{r} \equiv \int \tau_{KS}(\boldsymbol{r})d\boldsymbol{r}$$

(1.1)

where $\psi_i(r)$ are orbitals and $N_{el}$ is the number of electrons, or kinetic energy density (KED) $\tau_{KS}(r)$ or its positively-definite version[45,48,51,61,62]

$$\tau_+(r) = \frac{1}{2}\sum_{i=1}^{N_{el}}|\nabla\psi_i(r)|^2 = \tau_{KS}(r) + \frac{1}{4}\Delta\rho(r)$$

(1.2)

that integrates to the same $E_{kin}^{KS}$, as a function of density-dependent variables (features) that typically include different powers and orders of derivatives of the density. We neglect spin and partial occupancies without loss of generality and use atomic units unless stated otherwise.

In our previous works,[43,53] we showed that terms of the forth order gradient expansion can serve as effective descriptors – inputs for a machine learning algorithm when machine-learning the KED. The fourth order gradient expansion of the kinetic energy density is given by[63]

$$\tau_{GE4} = \tau_{TF}\left(1 + \frac{5}{27}p + \frac{20}{9}q + \frac{8}{81}q^2 - \frac{1}{9}pq + \frac{8}{243}p^2\right)$$

(1.3)

where $\tau_{TF}$ is the Thomas-Fermi[64] KED, $p = \frac{|\nabla\rho|^2}{4(3\pi^2)^{2/3}\rho^{8/3}}$ is the scaled squared gradient and $q = \frac{\Delta\rho}{4(3\pi^2)^{2/3}\rho^{5/3}}$ the scaled Laplacian of the density. The use of scaled gradient and Laplacian is also advantageous as it helps satisfy the scaling relations that can serve as useful constraints.[65] We also showed that another useful feature is the product of the density and Kohn-Sham effective potential $\rho V_{eff}$.[43,53] $V_{eff}$ is computable in OF-DFT unless a hybrid functional is used. Together with the terms of Eq. (1.3), this results in a set of features

$$x_\tau(r) = \left(\tau_{TF}(r), \tau_{TF}p(r), \tau_{TF}q(r), \tau_{TF}p^2(r), \tau_{TF}pq(r), \tau_{TF}q^2(r), \rho V_{eff}(r)\right)$$

(1.4)

A significant problem with the KED (rather than the KE) from the perspective of data-driven approaches such as ML are very uneven distributions of the target (the KED) and the features (see examples of respective data distributions in Refs. [43,53]). This complicates sampling and prima faci requires larger training sets, increasing computational cost. This is disadvantageous, in particular, with kernel methods that have to wield inverses of matrices of size $M \times M$ where $M$ is the number of training points.[59,60] Non-uniform sampling approaches or the use of committees can be used to palliate this issue[42,43] but imply additional complications of the method.

In Ref. [53], we showed that the issue of the very uneven data distributions can be effectively addressed by smoothing of both the features and the target on length scales sufficiently large to smoothen the distributions and simplify sampling while sufficiently small (e.g. smaller than the unit cell of crystalline materials) to preserve spatial structure and therefore the KEDF (kinetic energy density functional) ansatz. As long as smoothing is done with a linear operation, the smoothed KED integrates to the same KE, even though the overall effect of the smoothing of both the features and the target is not linear because the features are inputs to a non-linear method (Gaussian process regression, GPR,[60] in the case of Ref. [53]). With smoothing, when machine-learning from rather small training sets sampled with plain uniform sampling, accurate enough KE models of Al, Mg, and Si (simultaneously) could be built in Ref. [53] to result in errors on the order of 1% in a measure of the energy-volume dependence

$$B' = V_0^2 \frac{d^2 E}{dV^2} \approx \frac{E(V_0 - \Delta V) - 2E(V_0) + E(V_0 + \Delta V)}{(\Delta V/V_0)^2}$$

(1.5)

where $V_0$ is the equilibrium volume of the simulation cell and $E$ is the total energy.

When machine-learning KED, one works with large datasets such as the values of the features and the target on the entire Fourier grid of a plane wave method. The large raw dataset sizes pose challenges but also allow training a model on a small number of structures, even on a single material.[42] Features of Eq. (1.4) could potentially be fully averaged over the simulation cell; this would remove the need to work with large datasets but also evacuate much information. A single datapoint would result for one structure (material). If the number of structures on which the model is trained is low, this is problematic, and one could stay instead within the KEDF paradigm. But if the number of structures is high, then this may not be a problem. Moreover, averaging out spatial structure altogether would be advantageous as concatenation of KED data of many structures to learn a KE model which is portable across a large number of materials would quickly lead to potentially unwieldy datasets. For example, when using unit cells only of three materials only (Al, Mg, Si) with only three volumes per material (equilibrium volume, and uniformly compressed and expanded by $\Delta V$), the combined dataset contained about 580,000 points on a Fourier grid with a moderate plane wave cutoff of 500 eV. [43,53] Many millions of datapoints would have to be handled under less basic conditions (more materials, more structures per material, larger simulation cells, denser grids).

In this work, we show that when building a KE model from and for a large number of structures, fully averaging the features of Eq. (1.4) over the simulation cell, learning unit cell-averaged KED $\overline{\tau_{KS}}$ and computing $E_{kin}$ as a product of $\overline{\tau_{KS}}$ and cell volume is effective and results in computationally inexpensive and accurate models. The use of the cell-averaged values (of the target $\overline{\tau_{KS}}$ and features) rather than total kinetic energies palliates the issue of different cell volumes of different materials. We train a GPR model on multiple phases of unry, binary, and ternary compounds containing Li, Al, Mg, Si, As, Ga, Sb, Sn, Na, P and In, for a total of 433 materials. In addition to equilibrium structures, multiple strained structures are used for each material, for a total of 7794 structures. The model reproduces energy-volume *curves* (as opposed to only the values of *B'* explored in previous work[43,53]) around equilibrium geometry with good accuracy.

## 2 Methods

We developed a computational pipeline for a simple high-throughput setup in which structural information was obtained for different crystal phasis from the Materials Project via API interface[66] and converted in the Abinit[67,68] input files to perform Kohn-Sham DFT calculations on the equilibrium geometry as well as isotropically strained lattice. Information on electronic density, its gradient and Laplacian is output.

While details on the implementation will be presented elsewhere, we want to highlight the key points for the different stages of the pipeline. First, we choose a set of *N* elements with available local pseudopotentials, here Li, Al, Mg, Si, As, Ga, Sb, Sn, Na, P and In. All local pseudo potentials except for Na and Sn were taken from the Carter group repository.[21,27,28] Na was represented by the local pseudopotential developed by Legrain and Manzhos,[20] whereas the Sn local pseudopotentials was recently developed by us.[55] Then, we created combinations of elements from 1 to *N* without repetitions to define the chemical search space of materials. We restrict the search space of materials to crystals in conventional standard representation with 1 to $4 + 2m$ atoms in the basis where $m = \{1,2,3\}$ represents the number of elements in each combination of elements. In other words, we search for materials with stochiometric equivalent of $A_aB_bC_c$ with $1 \leq a + b + c \leq (4 + 2m)$ and *A, B*, and *C* are elements. Here, we restrict the database to materials with one to three different elements as well as to unit cells with unit cell vector not exceeding 1 nm in length. In addition, we apply an upper bound criterion for the energy above convex hall values given in the Materials Project of 400 meV and a lower bound criterion on the electronic density. The critical value was set to 90% of the minimal density (over the unit cell) computed for body-centered cubic lithium phases with the computational details stated below. This procedure yielded 433 materials.

The selected structures were then converted to a Abinit input format to perform DFT calculations. Any component of a lattice vector of less than $1 \times 10^{-9}$ nm was set to zero. The energy cutoff of the kinetic energy was set to 500 eV and the total energy was computed using the PBE exchange-correlation functional.[69,70] The *k*-point mesh was sampled at *k*-point densities of about $2.0 \times 10^{-3}$ nm$^3$ for most elements. Only for phases containing Li and Na atoms, we double the *k*-point density if the computed band gap on the given *k*-point grid was larger than 400 meV. The increased *k*-point density can increase accuracy of the total DFT energy especially for metallic phases. All structures were converted into their conventional standard unit cell representations. For each crystal phase, we performed volume optimization, i.e., the relative length and orientation of lattice vectors are maintained. The self-consistent field (SCF) calculations were converged with total energy differences of less than $2.7 \times 10^{-9}$ eV in two consecutive SCF cycles and the simulation cell volume until the isotropic stress is less than 0.7 kbar.

With the computed ground state geometry of the unit cell, we isotopically strained each structure with a strain $\epsilon$ between 0.93 and 1.1 in increments of 0.01, resulting in 18 data points for each material and a total of 7,794 structures. We recorded the computed DFT total energy and the kinetic energy, electron density, its gradient and Laplacian, as well as its kinetic energy density for each data point. The latter can be integrated to the kinetic energy and compared to the DFT computed kinetic energy. We ensure that the absolute difference of these values agrees within 0.1 mHa in our dataset for each data point.

The electron density, its gradient and Laplacian were output from Abinit and used to compute the features of Eq. (1.4) (except the term $\tau_{TF}q$ that integrated to zero and is therefore not used). They were averaged over the simulation cells and served as inputs to a GPR model. The features are thus

$$x = \left(\overline{\tau_{TF}}, \overline{\tau_{TF}p}, \overline{\tau_{TF}p^2}, \overline{\tau_{TF}pq}, \overline{\tau_{TF}q^2}, \overline{\rho V_{eff}}\right)$$

(2.1)

where $\overline{\phantom{x}}$ (bar) indicates spatial averaging. The kinetic energy output from Abinit and averaged over the simulation cell, $\overline{\tau_{KS}}$, served as the target. The averaging of the features and target (rather than using integrated values) is useful to palliate the effect of differences in simulation cell sizes. The kinetic energy is then obtained by multiplying the averaged KED predicted by GPR by the volume of the unit cell. GPR was performed in Matlab using *fitrgp* function. The features were scaled on the unit cube, and an isotropic Matern52 kernel was used, which provided slightly better results than RBF and Matern32 kernels. The reader is referred to the literature for the description of the GPR method and Matern kernels.[60,71] The length and regularization (noise) hyperparameters were optimized by scanning and monitoring the error on the test set. The optimal length parameter was $l = 7$ and the noise parameter $\sigma = 1\times10^{-4}$. 20% of the data (uniformly and randomly drawn from the full set) were reserved for testing unless otherwise indicated (e.g. when training on monoelemental and testing on multielemental compounds). The final GPR model available to the readers was trained on all data using the hyperparameters determined with the help of the test set.

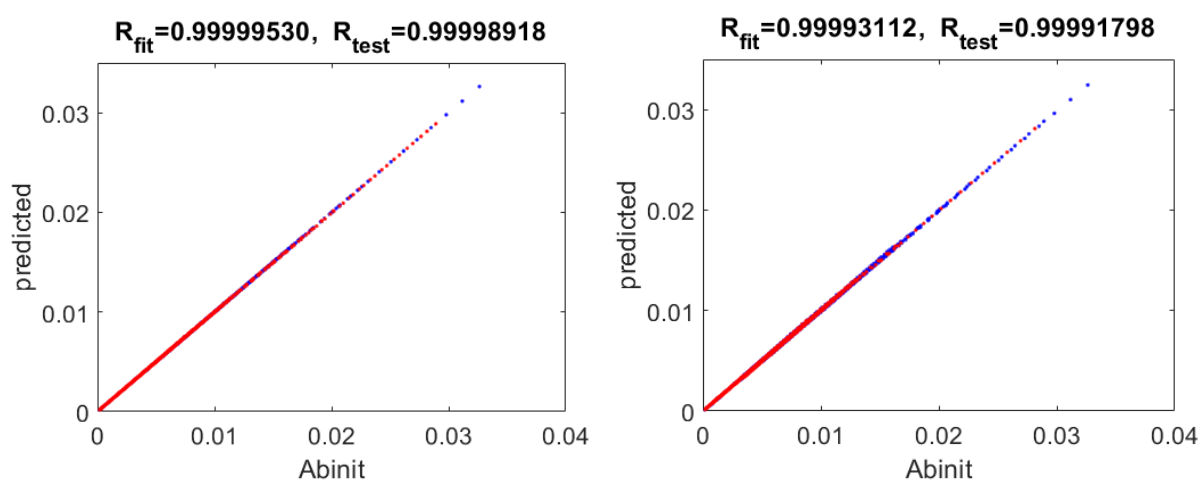

Figure 1. Correlation plots between the reference Abinit-based $\overline{\tau_{KS}}$ and those predicted by the GPR model (left) and linear regression (right). Blue: train points, red: test points. The correlation coefficients for train ($R_{fit}$) and test ($R_{test}$) points are shown above the plots.

## 3  Results

### 3.1  GPR model

We first present the results of GPR on the entire dataset, whereby train and test materials were chosen uniformly randomly from the entire dataset combining unary, binary and ternary compounds. The fit quality of $\overline{\tau_{KS}}$ by the GPR model is very high, the correlation plots between the reference Abinit values and the model are shown in Figure 1, left panel. The correlation coefficients for both training and test sets are 1 to within 5 digits. The resulting energy-volume curves for unary, binary and ternary compounds, respectively, are presented in Figure 2, Figure 3, and Figure 4, respectively. A subset of 60 binary compounds is shown in Figure 3, see SI for the plots for the other compounds. The total energy in these plots is computed by subtracting the DFT kinetic energy from DFT total energy and adding instead the kinetic energy predicted by the ML model. The blue dots are reference DFT data and red or green dots are values predicted by the model, whereby red corresponds to materials in the training set and green to those in the test set. Atom labels (chemical formulas) and Materialsproject ID numbers are shown on the respective graphs in black. The values of $B'$ (computed using $\Delta V/V_0 = 3\%$) are also shown on the graphs: the numbers in blue are from reference DFT calculation, and those in red are from energies predicted by the ML model. One can appreciate visually that the quality of the energy prediction for the test set is by and large similar to that for the train set.

Table 1. Statistics (mean, minimum, and maximum over 20 GP or polynomial regressions differing by random selections of training and test data) of the RMSE (root mean square errors) of averaged kinetic energy density $\overline{\tau_{KS}}$ as well as MRE (mean relative error) and MDRE (median relative error) in the curvature of the energy-volume curve $B'$. Results for linear regression are given without such statistics as they do not significantly vary in function of a particular train-test split.

|  | RMSE in $\overline{\tau_{KS}}$, $10^5$ a.u. | | | MRE in $B'$ | | | MDRE in $B'$ | | |
|---|---|---|---|---|---|---|---|---|---|
|  | mean | min | max | mean | min | max | mean | min | max |
|  | GPR | | | | | | | | |
| Train | 1.05 | 1.00 | 1.09 | 0.13 | 0.11 | 0.14 | 0.08 | 0.07 | 0.09 |
| Test | 1.47 | 1.13 | 1.84 | 0.13 | 0.09 | 0.16 | 0.08 | 0.06 | 0.10 |
|  | Linear regression | | | | | | | | |
| Train | 4.14 | | | 0.57 | | | 0.24 | | |
| Test | 3.98 | | | 0.53 | | | 0.25 | | |
|  | Polynomial regression, $\Pi = 4$ | | | | | | | | |
| Train | 1.01 | 0.95 | 1.06 | 0.13 | 0.11 | 0.17 | 0.09 | 0.08 | 0.10 |
| Test | 2.69 | 1.30 | 9.4 | 0.12 | 0.10 | 0.17 | 0.09 | 0.07 | 0.11 |
|  | Polynomial regression, $\Pi = 5$ | | | | | | | | |
| Train | 0.89 | 0.82 | 0.95 | 0.14 | 0.11 | 0.37 | 0.09 | 0.07 | 0.10 |
| Test | 6.48 | 1.03 | 20.35 | 0.12 | 0.09 | 0.18 | 0.09 | 0.07 | 0.11 |

The statistics over 20 runs differing by such selection is given in Table 1. We provide MDRE in addition to MRE because the latter is easily influenced by a small number of materials with high errors. In all cases, the GPR model results in a pronounced minimum on the energy-volume curve, which is important for structure optimization. All figures here and below are obtained with the same test-train split (obtained by fixing the random seed), to facilitate comparison.

*3.2 Advantage of GPR over linear and polynomial regression*

To highlight the value of machine learning with GPR, we also present results with linear and polynomial regressions. These regressions can be obtained by using dot product kernels of the form $k(x, x') = \sum_{p=1}^{\Pi}(xx')^p$,[59,72,73] where $\Pi$ defines the order of the polynomial. We first show the results of a linear model in Figure 5, Figure 6, and Figure 7 for unary, binary (same subset as in Figure 3) and ternary compounds, respectively. The results are also summarized in Table 1 and show much worse results and in terms of predicting the kinetic energy and in terms of estimating the curvature *B'* of the energy volume curve, which is off on average by dozens per cent, vs errors on the order of 10% for GPR. Many materials do not show a pronounced minimum of the curve. In this case, different train-test splits do not have a significant effect on the results; linear regression being more robust (showing similar train and test set errors) at the price of expressive power. The correlation plots between the reference Abinit values of $\overline{\tau_{KS}}$ and the linear model are shown in Figure 1, right panel. By this measure, the quality of the fit appears to be high even though the energy-volume curves are unimpressive: the correlation coefficients for both training and test sets are 1 to within 4 digits (while it is 1 to within 5 digits with GPR). This highlights that in this application, a very high quality of ML is required to result in practically useful machine-learned kinetic energy functionals.

The model accuracy can be improved by using a higher-order polynomial model. Even with $\Pi = 4$ the model is not quite as accurate as that with a Matern kernel, and with $\Pi = 5$ one observes significant overfitting (test errors higher than with linear regression). The results are summarized in Table 1. The graphs for the case of $\Pi = 4$ are shown in SI. The clear advantage of GPR with a Matern kernel over low-order polynomial regression also signifies that the space of materials is adequately sampled.[72]

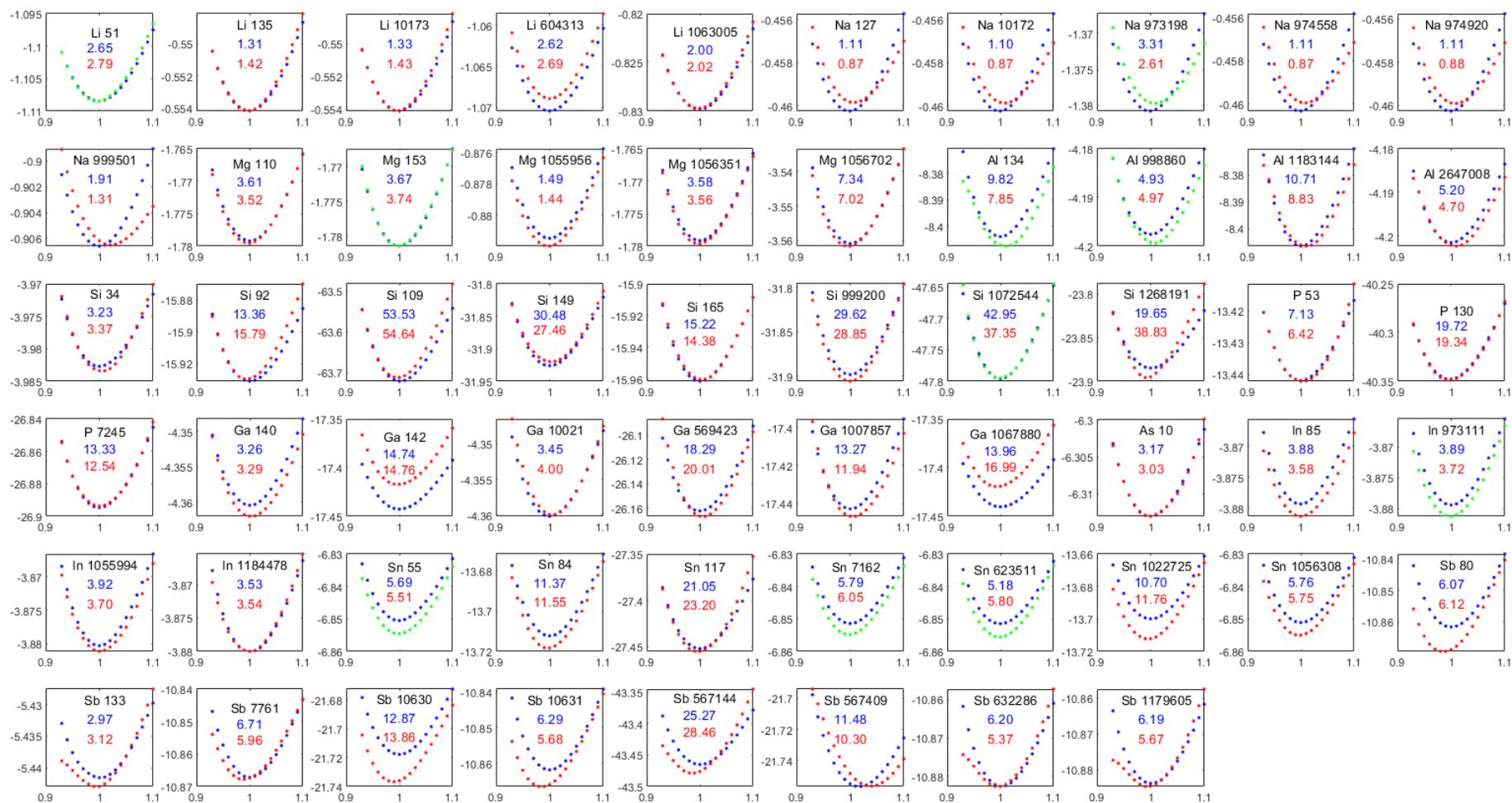

Figure 2. Energy-volume curves for unary compounds. The abscissa is relative strain $V/V_0$ (where $V_0$ is equilibrium volume) and the ordinate axis is energy in *a.u*. The blue dots are reference DFT data and red (training set) and green (test set) dots are values predicted by the GPR model. Atom labels and Materialsproject ID numbers are shown on the graphs in black. Numbers in blue are the values of $B'$ from reference DFT calculation, and those in red are the values of $B'$ predicted by the ML model.

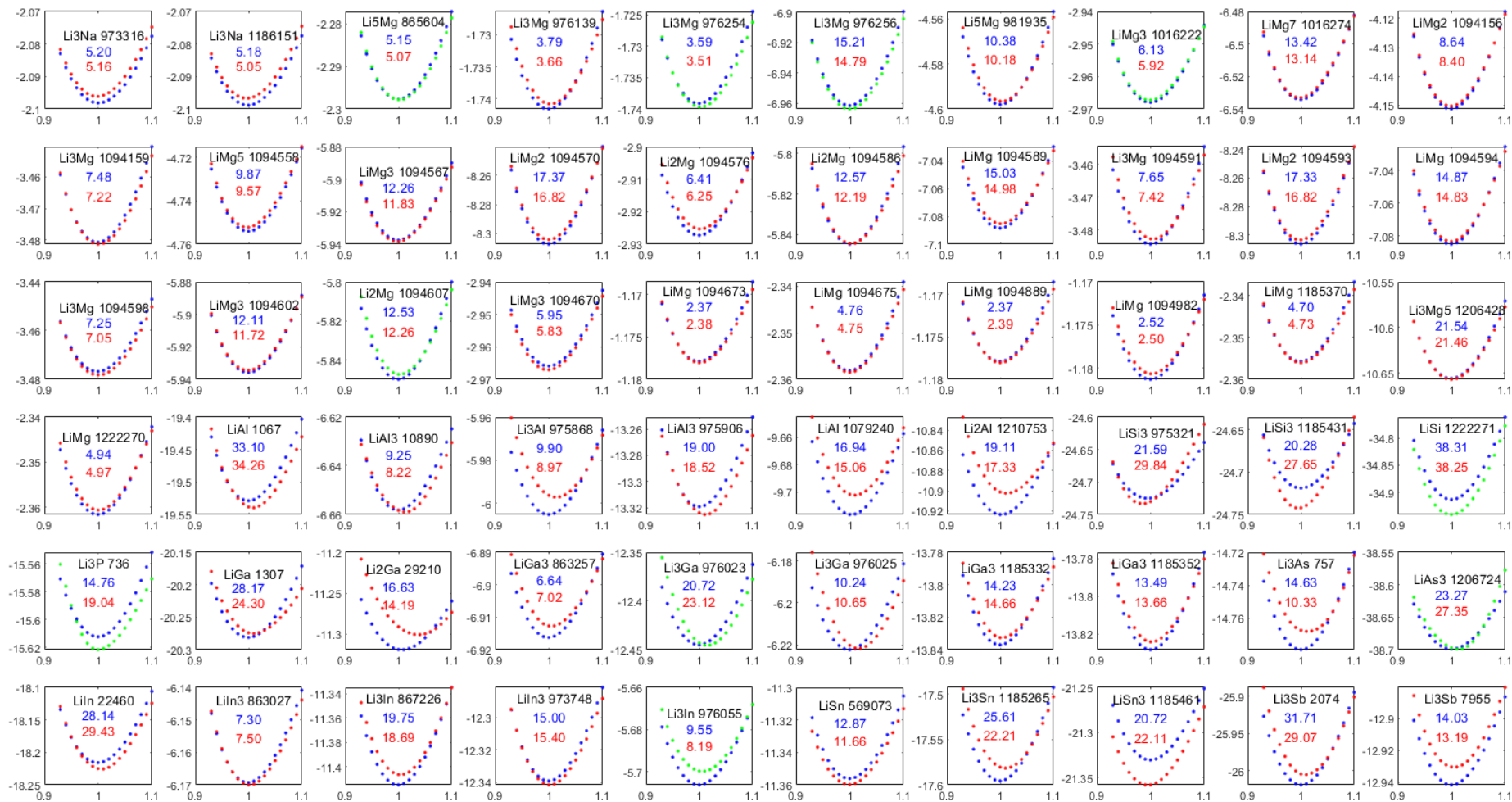

Figure 3. Energy-volume curves for a subset of binary compounds (the curves for other compounds are shown in SI). The abscissa is relative strain $V/V_0$ (where $V_0$ is equilibrium volume) and the ordinate axis is energy in *a.u*. The blue dots are reference DFT data and red (training set) and green (test set) dots are values predicted by the GPR model. Atom labels and Materialsproject ID numbers are shown on the graphs in black. Numbers in blue are the values of *B*' from reference DFT calculation, and those in red are the values of *B*' predicted by the ML model.

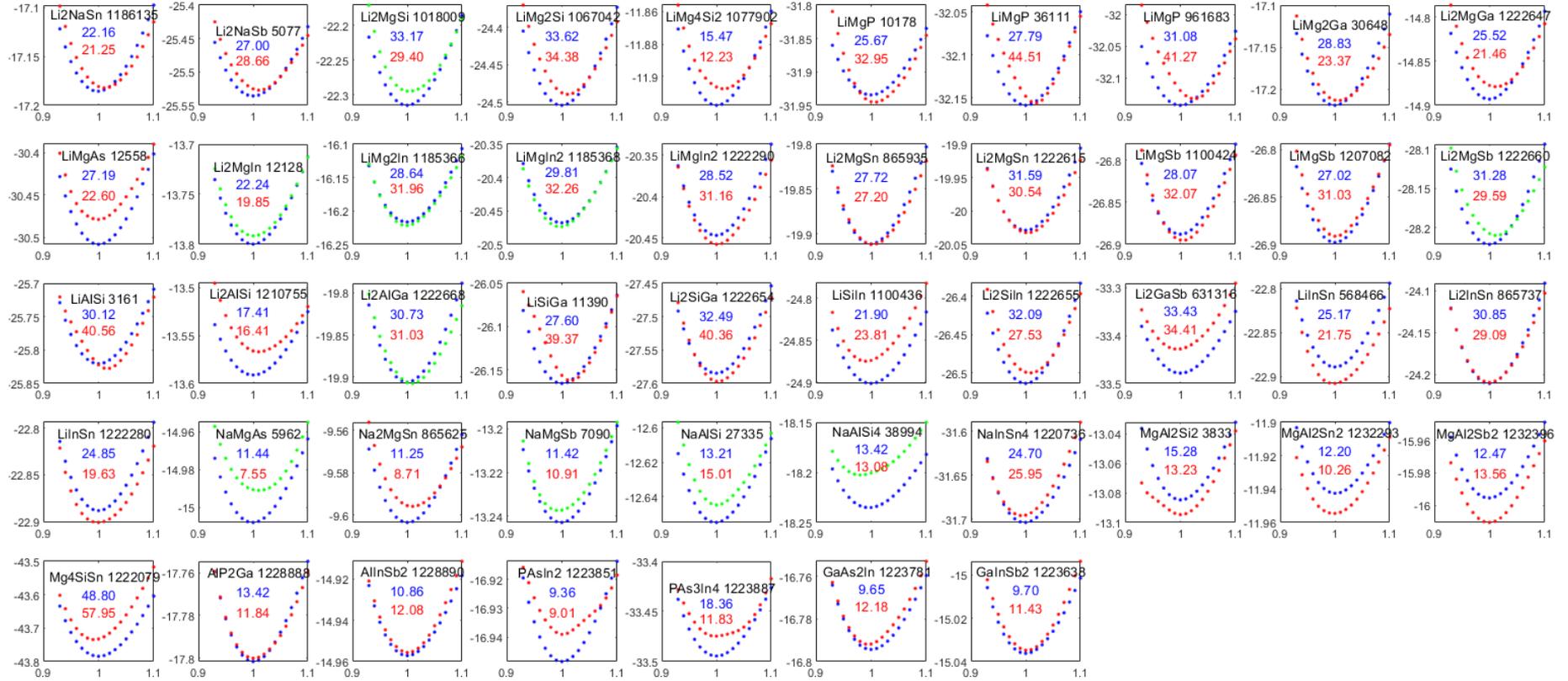

Figure 4. Energy-volume curves for ternary compounds. The abscissa is relative strain $V/V_0$ (where $V_0$ is equilibrium volume) and the ordinate axis is energy in *a.u*. The blue dots are reference DFT data and red (training set) and green (test set) dots are values predicted by the GPR model. Atom labels and Materialsproject ID numbers are shown on the graphs in black. Numbers in blue are the values of $B'$ from reference DFT calculation, and those in red are the values of $B'$ predicted by the ML model.

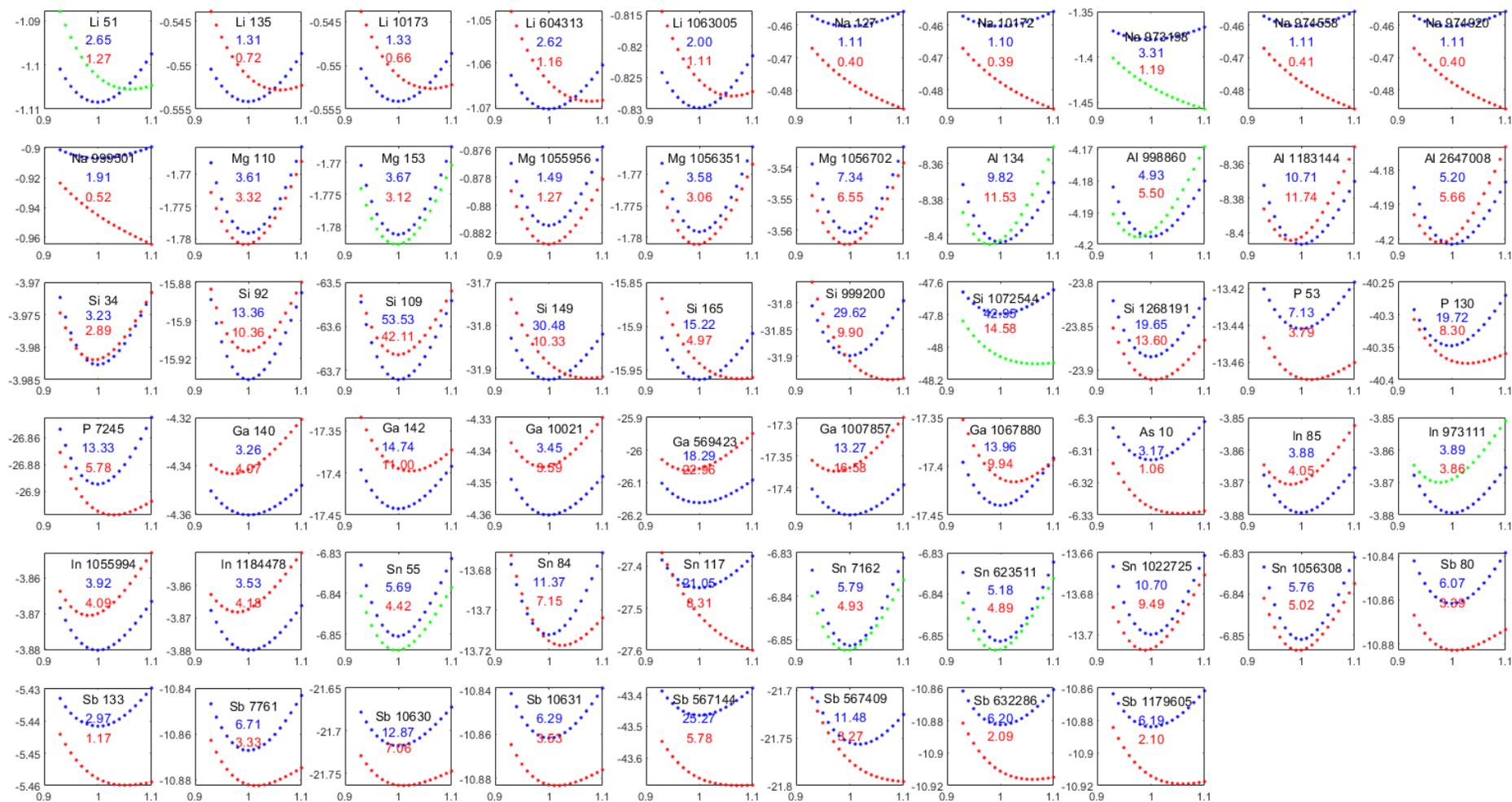

Figure 5. Energy-volume curves for unary compounds. The abscissa is relative strain $V/V_0$ (where $V_0$ is equilibrium volume) and the ordinate axis is energy in *a.u.* The blue dots are reference DFT data and red (training set) and green (test set) dots are values predicted by linear regression. Atom labels and Materialsproject ID numbers are shown on the graphs in black. Numbers in blue are the values of $B'$ from reference DFT calculation, and those in red are the values of $B'$ predicted by the model.

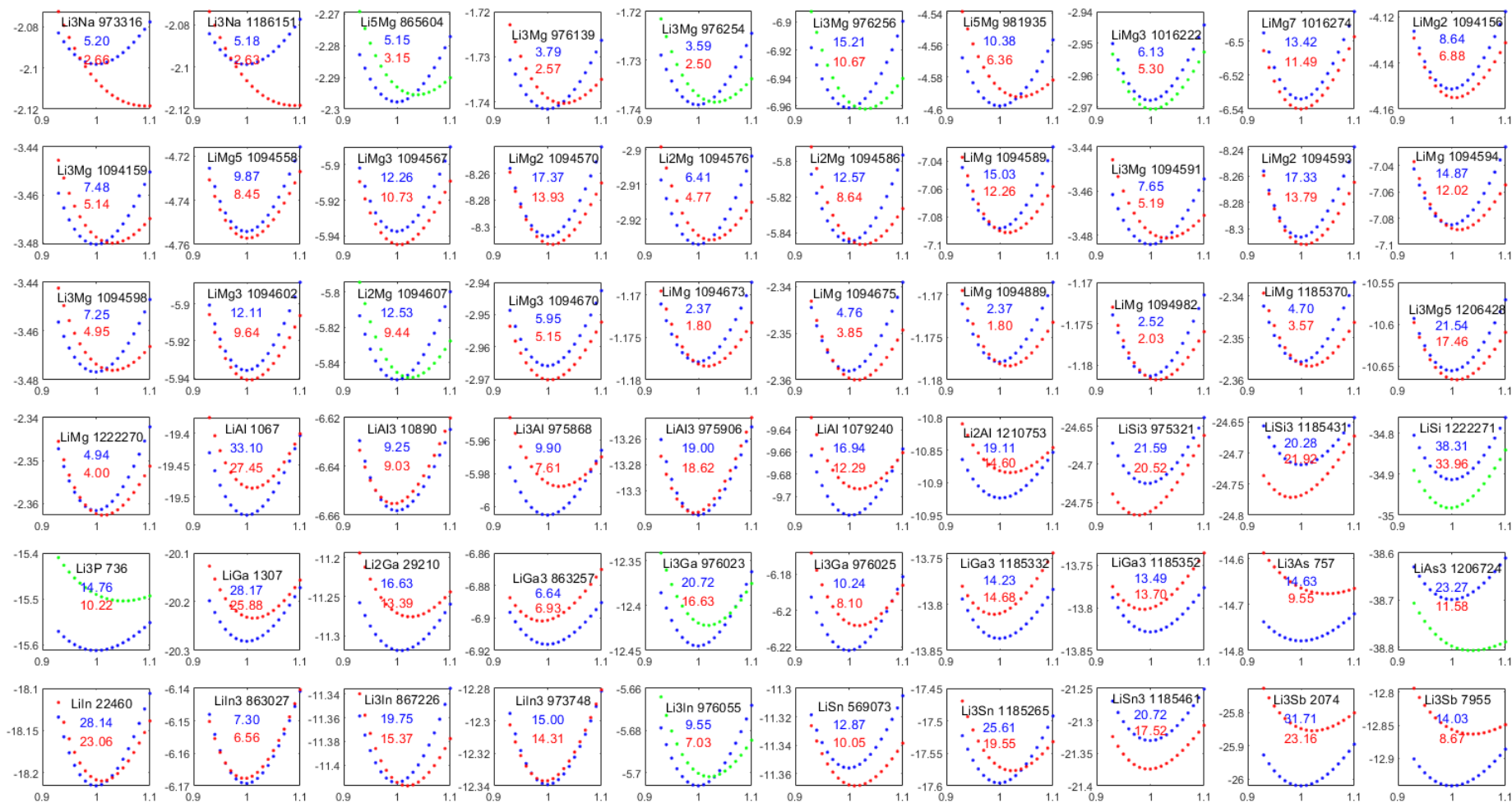

Figure 6. Energy-volume curves for the same subset of binary compounds as in Figure 3. The abscissa is relative strain $V/V_0$ (where $V_0$ is equilibrium volume) and the ordinate axis is energy in *a.u.* The blue dots are reference DFT data and red (training set) and green (test set) dots are values predicted by linear regression. Atom labels and Materialsproject ID numbers are shown on the graphs in black. Numbers in blue are the values of $B'$ from reference DFT calculation, and those in red are the values of $B'$ predicted by the model.

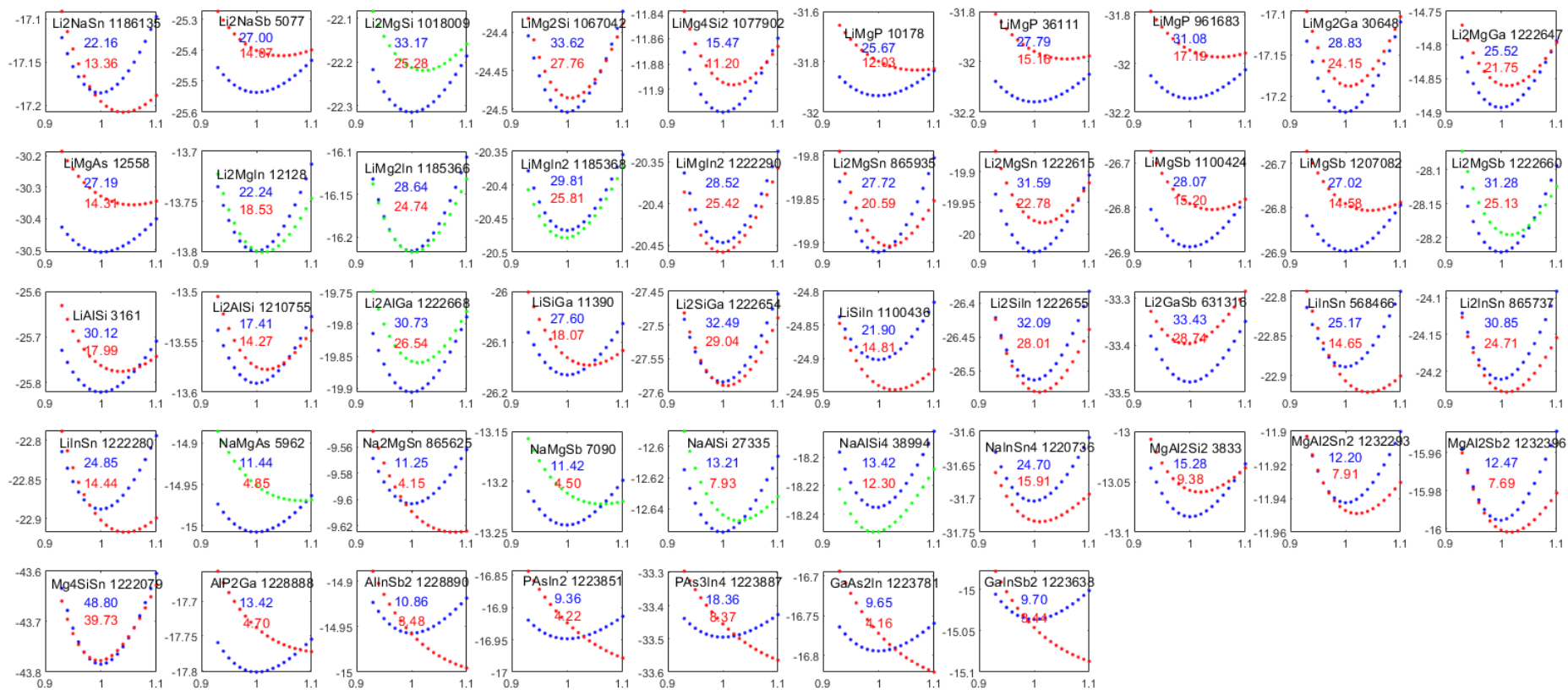

Figure 7. Energy-volume curves for ternary compounds. The abscissa is relative strain $V/V_0$ (where $V_0$ is equilibrium volume) and the ordinate axis is energy in *a.u.* The blue dots are reference DFT data and red (training set) and green (test set) dots are values predicted by linear regression. Atom labels and Materialsproject ID numbers are shown on the graphs in black. Numbers in blue are the values of $B'$ from reference DFT calculation, and those in red are the values of $B'$ predicted by the model.

*3.3    Portability among unary, binary, and ternary compounds*

To test portability beyond (necessarily relatively small) test sets, we train the GPR model only on unary compounds and test it on binary and ternary compounds. The results are shown in Figure 8, Figure 9 (same subset as in Figure 3), and Figure 10. The train / test rmse of $\overline{\tau_{KS}}$ is 3.86×10$^{-6}$ / 3.29×10$^{-5}$ a.u., and the mean (median) relative error in *B'* is 0.10 (0.08) / 0.23 (0.16) (cf. Table 1). The drop in accuracy vs the model of section 3.1 is not surprising given the small number of unary materials, i.e. only 58. What is encouraging, however, is that *all* binary and ternary compounds still show pronounced minima on the energy-volume curves.

Training on only binary compounds (a training set of 328 materials out of 433, i.e. 76% and close to the 80/20 split used when fitting all types of materials) results in a model that can reproduce the ternary compounds relatively well but not necessarily the unary compounds. The results are shown in Figure 11, Figure 12, and Figure 13. Train / test rmse of $\overline{\tau_{KS}}$ is 1.01×10$^{-5}$ / 3.23×10$^{-4}$ a.u., and the mean (median) relative error in *B'* is 0.15 (0.08) / 0.14 (0.11) (cf. Table 1). Interestingly, while extrapolation ability to ternary compounds is decent, extrapolation to unary compounds is relatively poor. This has to do with the distribution of the data. The distributions of features and targets for unary, binary, and ternary compounds are shown in Figure 14. One observes, somewhat counterintuitively, that unary compounds cover a wider range of feature and target values than binary and ternary ones. Especially the range of values close to 0 is sampled well by unary but not multielemental compounds. It is therefore important to include monoelemental materials in the training set.

*3.4    Final model*

The final GPR model available to interested readers was trained on all data. It resulted in a rmse of $\overline{\tau_{KS}}$ of 1.06×10$^{-5}$ a.u., and the mean (median) relative error in *B'* of 0.13 (0.07). The model is available at https://github.com/sergeimanzhos/GPRKE in Matlab format in the form of matrices $c$ and $x_{train}$, whereby the output $\overline{\tau_{KS}}(x) = \sum_{i=1}^{N} c_i k(x, x_{train}^{(i)})$ and *i* runs over the training points, as well as the definition of the kernel function $k(x, x')$. A Matlab GPR object obtained with *fitrgp* function is also available; as it is about 236 MB in size, it can be obtained from the authors at reasonable request. The model provided on Github is a manual re-coding of GPR equations that results in the same output as the model obtained with Matlab's *fitrgp* function while allowing model export with much smaller data size (about 400 KB). Sharing in the form of $c$ and $x_{train}$ also facilitates porting the model to other programming environments if needed.

When using the model, the user is advised to form the features of $x$ of Eq. (2.1) (computed by the user for materials of interest) using a DFT code of their choice, minmax scale them on unit cube with values provided with the model, call the GPR model to obtain averaged KED, and obtain the kinetic energy by multiplying the average KED by the volume of the cell.

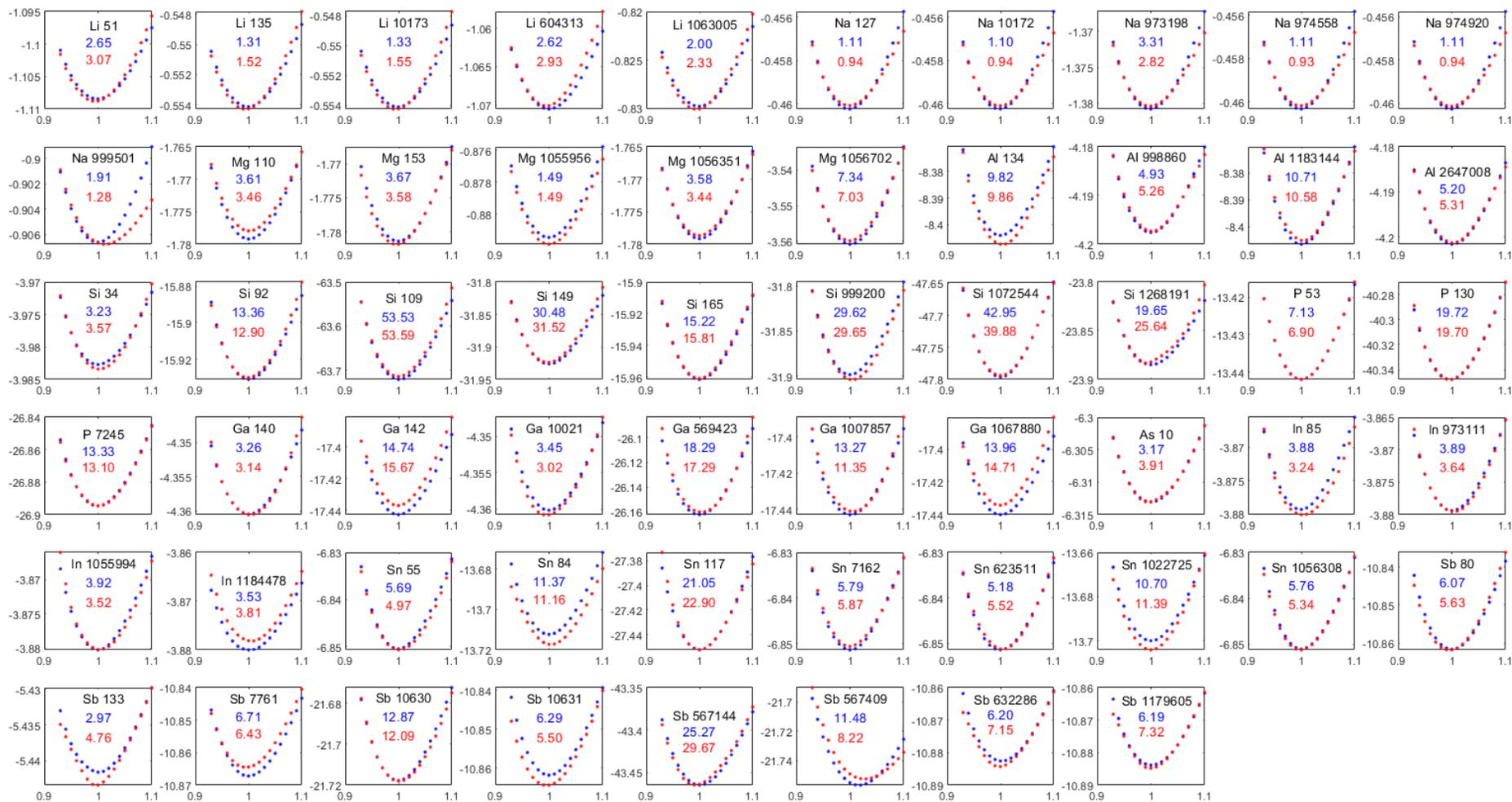

Figure 8. Energy-volume curves for unary compounds when training a GPR model on unary compounds. The abscissa is relative strain $V/V_0$ (where $V_0$ is equilibrium volume) and the ordinate axis is energy in *a.u*. The blue dots are reference DFT data and red dots are values predicted by the GPR model. Atom labels and Materialsproject ID numbers are shown on the graphs in black. Numbers in blue are the values of $B'$ from reference DFT calculation, and those in red are the values of $B'$ predicted by the ML model.

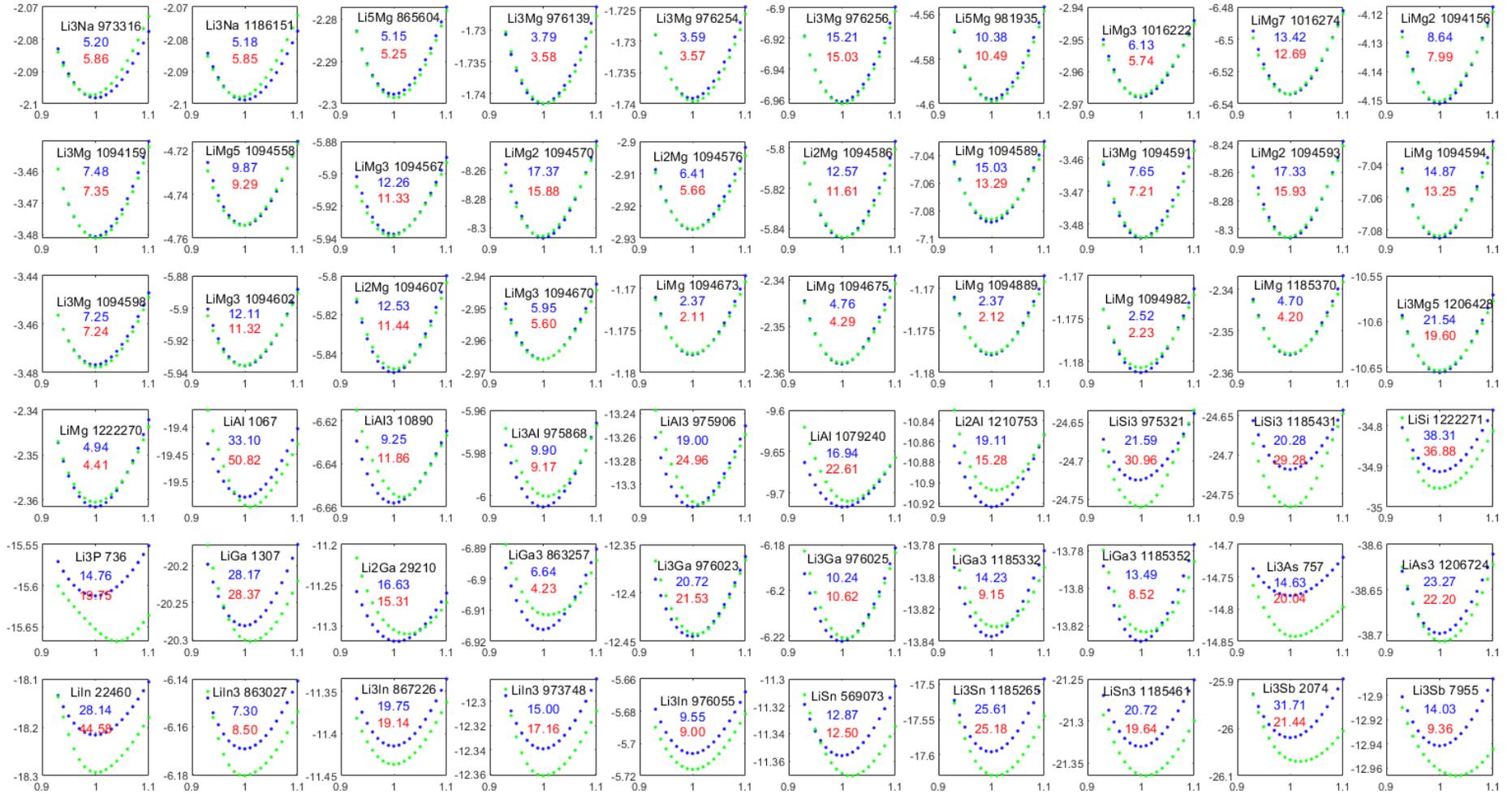

Figure 9. Energy-volume curves for binary compounds (for the same subset of as in Figure 3) when training a GPR model on unary compounds. The abscissa is relative strain $V/V_0$ (where $V_0$ is equilibrium volume) and the ordinate axis is energy in *a.u*. The blue dots are reference DFT data and green dots are values predicted by the GPR model. Atom labels and Materialsproject ID numbers are shown on the graphs in black. Numbers in blue are the values of $B'$ from reference DFT calculation, and those in red are the values of $B'$ predicted by the ML model.

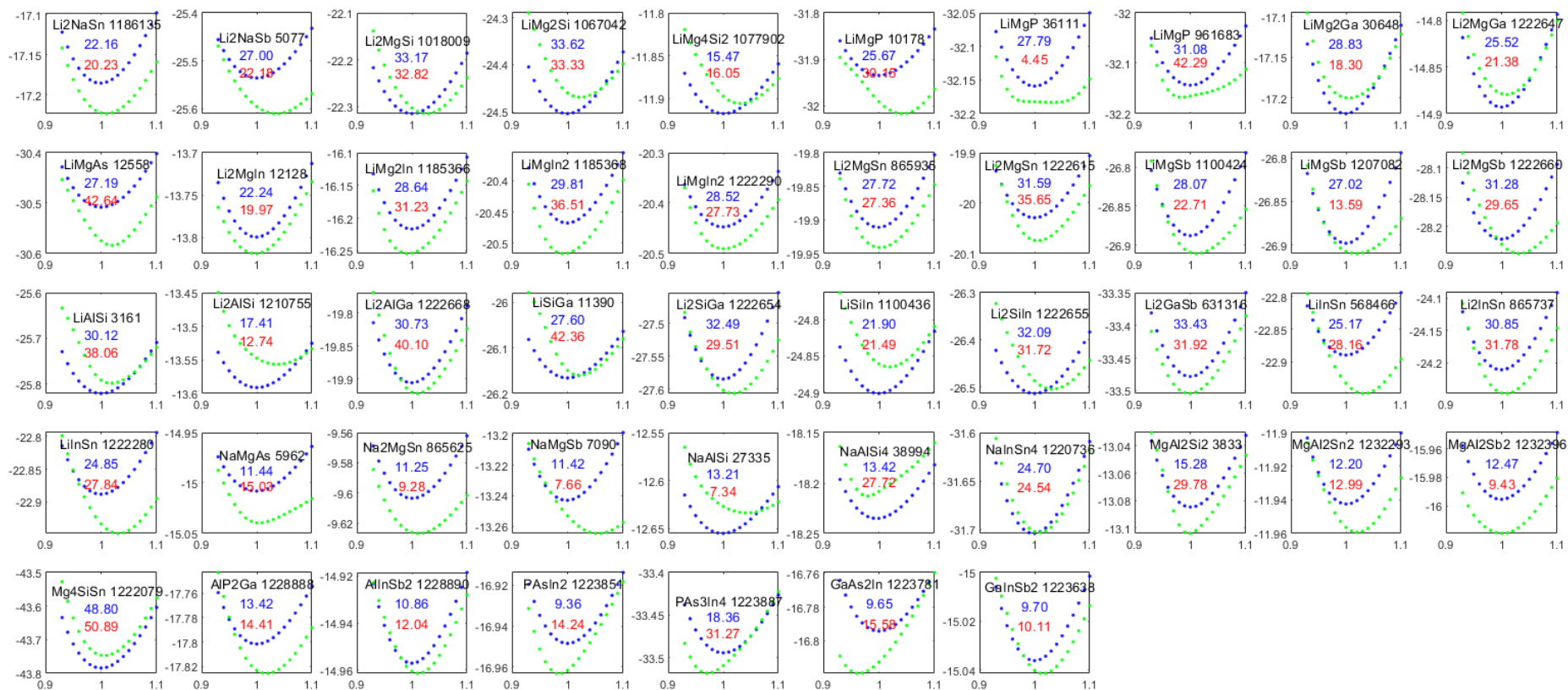

Figure 10. Energy-volume curves for ternary compounds when training a GPR model on unary compounds. The abscissa is relative strain $V/V_0$ (where $V_0$ is equilibrium volume) and the ordinate axis is energy in *a.u*. The blue dots are reference DFT data and green dots are values predicted by the GPR model. Atom labels and Materialsproject ID numbers are shown on the graphs in black. Numbers in blue are the values of *B'* from reference DFT calculation, and those in red are the values of *B'* predicted by the ML model.

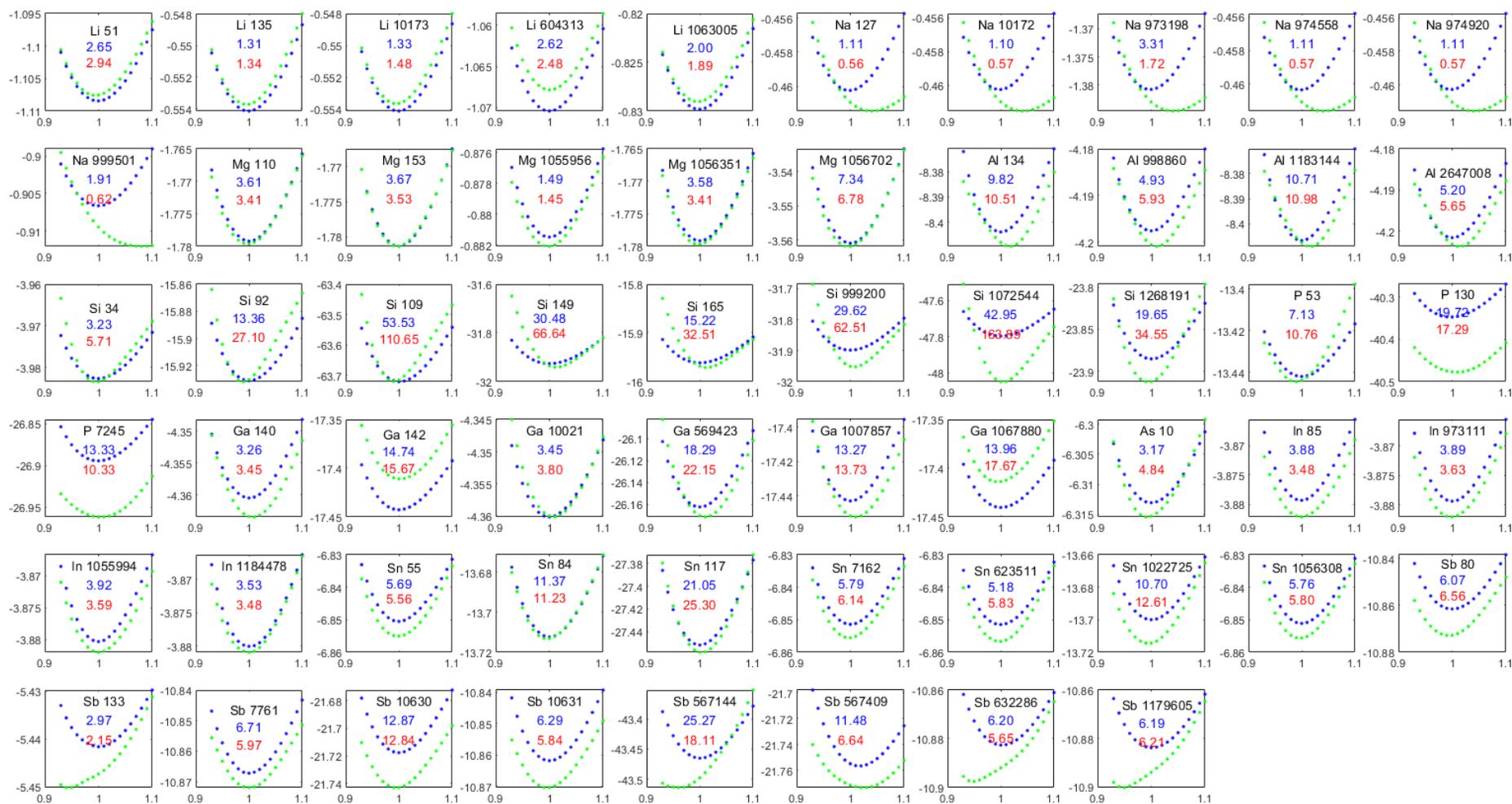

Figure 11. Energy-volume curves for unary compounds when training a GPR model on binary compounds. The abscissa is relative strain $V/V_0$ (where $V_0$ is equilibrium volume) and the ordinate axis is energy in $a.u$. The blue dots are reference DFT data and green dots are values predicted by the GPR model. Atom labels and Materialsproject ID numbers are shown on the graphs in black. Numbers in blue are the values of $B'$ from reference DFT calculation, and those in red are the values of $B'$ predicted by the ML model.

Figure 12. Energy-volume curves for binary compounds (for the same subset of as in Figure 3) when training a GPR model on binary compounds. The abscissa is relative strain $V/V_0$ (where $V_0$ is equilibrium volume) and the ordinate axis is energy in *a.u.* The blue dots are reference DFT data and red dots are values predicted by the GPR model. Atom labels and Materialsproject ID numbers are shown on the graphs in black. Numbers in blue are the values of $B'$ from reference DFT calculation, and those in red are the values of $B'$ predicted by the ML model.

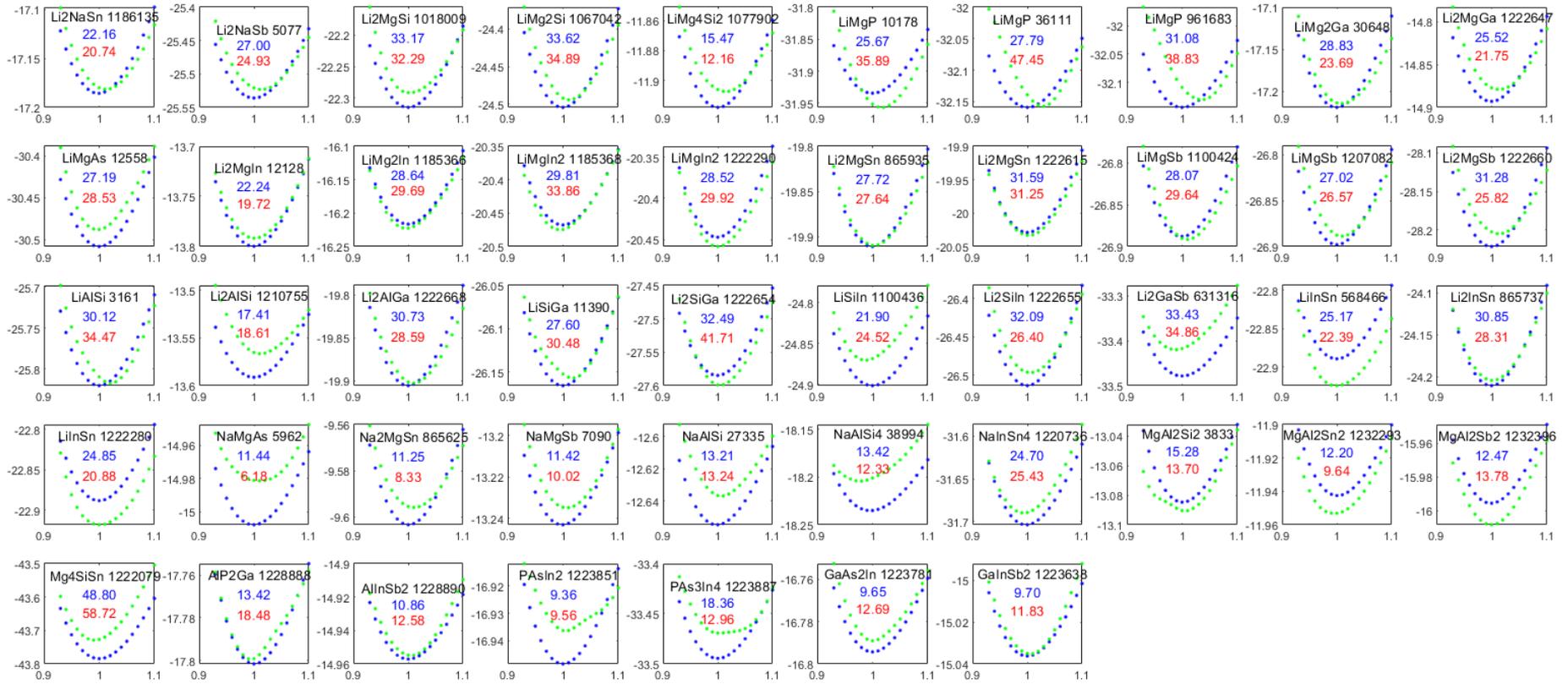

Figure 13. Energy-volume curves for ternary compounds when training a GPR model on binary compounds. The abscissa is relative strain $V/V_0$ (where $V_0$ is equilibrium volume) and the ordinate axis is energy in *a.u.* The blue dots are reference DFT data and green dots are values predicted by the GPR model. Atom labels and Materialsproject ID numbers are shown on the graphs in black. Numbers in blue are the values of $B'$ from reference DFT calculation, and those in red are the values of $B'$ predicted by the ML model.

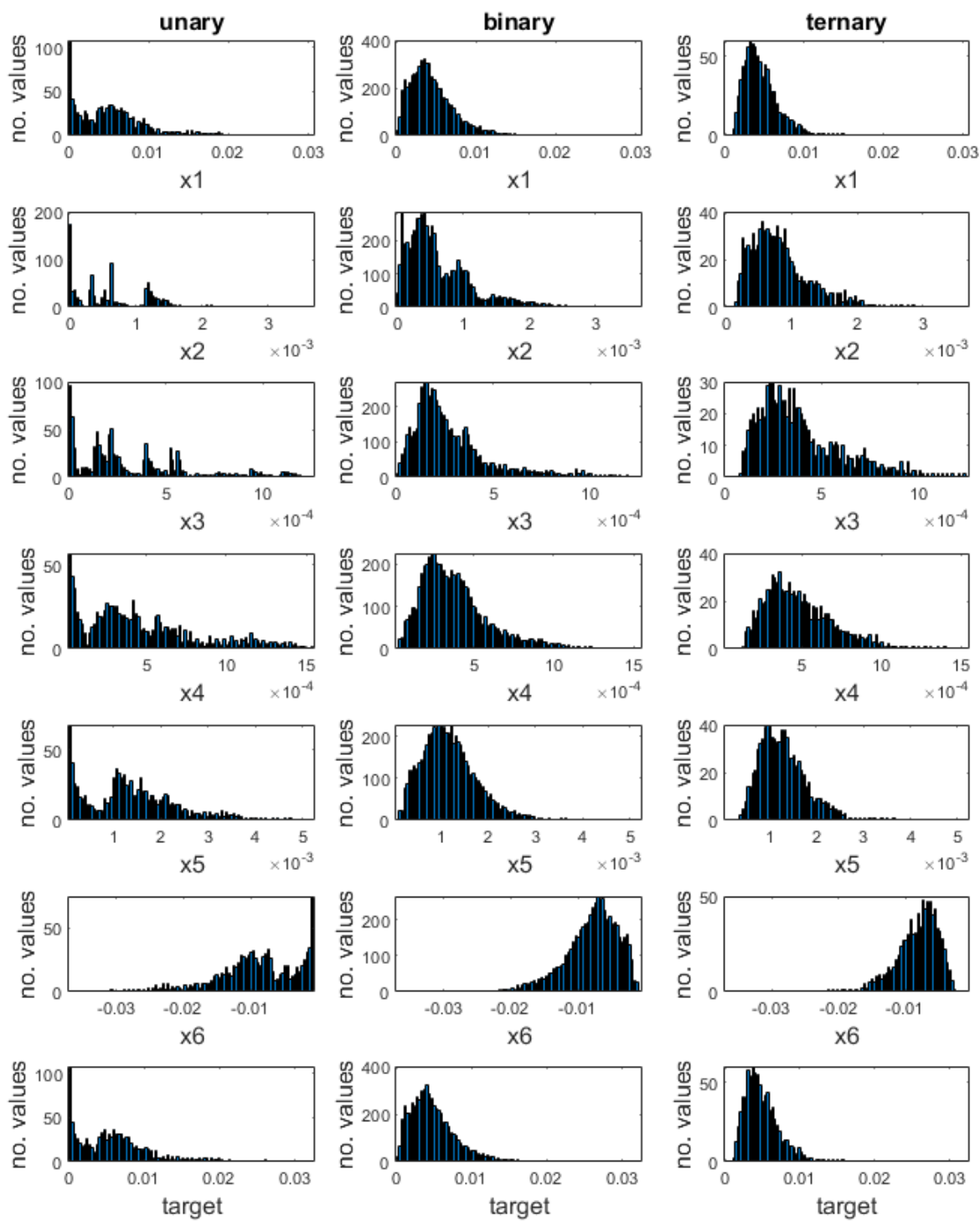

Figure 14. Distributions (histograms) of feature and target values for unary, binary, and ternary compounds, where "x1" to "x6" are the components of $\boldsymbol{x}$ from Eq. (2.1) (i.e. $\overline{\tau_{TF}}, \overline{\tau_{TF}p}, \overline{\tau_{TF}p^2}, \overline{\tau_{TF}pq}, \overline{\tau_{TF}q^2}, \overline{\rho V_{eff}}$) and the target is $\overline{\tau_{KS}}$.

# 4 Conclusions

We presented a machine-learned model of kinetic energy (KE) for orbital-free density functional theory suitable for bulk light weight metals and compounds made of group III-V elements. We used Gaussian process regression (GPR) to machine-learn the averaged over the unit cell kinetic energy density (KED) $\overline{\tau_{KS}}$. The features are terms of the 4th order gradient expansion averaged over the unit cell. The use of the cell-averaged values rather than integrated kinetic energies and features palliates the issue of different cell volumes of different materials, with kinetic energies, as an integral of the KED over the volume, being an extensive quantity.

The $\overline{\tau_{KS}}$ was machine-learned from data computed with Kohn-Sham DFT with plane wave bases and local pseudopotentials. The dataset included multiple phases of unary, binary, and ternary compounds containing Li, Al, Mg, Si, As, Ga, Sb, Na, Sn, P, and In. A total of 433 materials were used for training, and 18 strained structures were used for each material. Kinetic energy predicted by the model allows reproducing energy-volume curves around equilibrium geometry with good accuracy. We were able to achieve pronounced minima on energy-volume curves for all materials in train and test sets, with measures of curvature around the minimum within 9% of their KS DFT reference values for most materials.

We observed that a GPR model trained on a small number of unary compounds is able to extrapolate relatively well to binary and ternary compounds but not vice versa. This was related to the fact that unary compound sample a wider region of the descriptor space than binary and ternary compounds and it is therefore important to include them in the training set.

We showed that the GPR model is superior to linear and polynomial regressions, resulting in significantly better energy-volume curves. Even though there is a substantial difference in the quality of the KE model between GPR and simple linear and polynomial regressions, in all cases the correlation coefficients for both training and test sets are 1 to within at least 4 digits and the correlation plots – an often used way to evaluate the quality of ML in various applications - are practically linear. This highlights that in this particular application a very high quality of ML, beyond of what is typically required and achievable in other applications such as force fields or materials informatics, is required to result in practically useful machine-learned kinetic energy functionals.

# 5 Supplementary Material

Energy-volume curves for selected models and data subsets.

# 6  Acknowledgements

This work was supported by JST-Mirai Program Grant Number JPMJMI22H1, Japan. S. M. thanks Tucker Carrington for discussions. J. L. acknowledged financial support from the National Science and Technology Council (NSTC) under Grant Number 110-2112-M-110-025-MY3.

# 7  Conflicts of interest

The authors declare no conflicts of interest.

# 8  Data availability statement

The data as well as the GPR model are available at https://github.com/sergeimanzhos/GPRKE .

[13] K. Jiang, X. Shao, and M. Pavanello, "Efficient time-dependent orbital-free density functional theory: Semilocal adiabatic response," Phys. Rev. B **106**(11), 115153 (2022).

[14] D. Koch, M. Chaker, M. Ihara, and S. Manzhos, "Density-Based Descriptors of Redox Reactions Involving Transition Metal Compounds as a Reality-Anchored Framework: A Perspective," Molecules **26**(18), 5541 (2021).

[15] S.R. Kirk, and S. Jenkins, "Beyond energetic and scalar measures: Next generation quantum theory of atoms in molecules," WIREs Computational Molecular Science **12**(6), e1611 (2022).

[16] J.I. Rodríguez, F. Cortés-Guzmán, and J.S.M. Anderson, editors, *Advances in Quantum Chemical Topology Beyond QTAIM* (Elsevier, Amsterdam, 2023).

[17] M. Chen, X.-W. Jiang, H. Zhuang, L.-W. Wang, and E.A. Carter, "Petascale Orbital-Free Density Functional Theory Enabled by Small-Box Algorithms," J. Chem. Theory Comput. **12**(6), 2950–2963 (2016).

[18] D.J. González, and L.E. González, "Structure and motion at the liquid-vapor interface of some interalkali binary alloys: An orbital-free ab initio study," J. Chem. Phys. **130**(11), 114703 (2009).

[19] K.M. Carling, and E.A. Carter, "Orbital-free density functional theory calculations of the properties of Al, Mg and Al–Mg crystalline phases," Modelling Simul. Mater. Sci. Eng. **11**(3), 339 (2003).

[20] F. Legrain, and S. Manzhos, "Highly accurate local pseudopotentials of Li, Na, and Mg for orbital free density functional theory," Chem. Phys. Lett. **622**, 99–103 (2015).

[21] C. Huang, and E.A. Carter, "Transferable local pseudopotentials for magnesium, aluminum and silicon," Phys. Chem. Chem. Phys. **10**(47), 7109–7120 (2008).

[22] Q. Liu, D. Lu, and M. Chen, "Structure and dynamics of warm dense aluminum: a molecular dynamics study with density functional theory and deep potential," J. Phys.: Condens. Matter **32**(14), 144002 (2020).

[23] H. Zhuang, M. Chen, and E.A. Carter, "Elastic and Thermodynamic Properties of Complex Mg-Al Intermetallic Compounds via Orbital-Free Density Functional Theory," Phys. Rev. Applied **5**(6), 064021 (2016).

[24] J.-D. Chai, V.L. Lignères, G. Ho, E.A. Carter, and J.D. Weeks, "Orbital-free density functional theory: Linear scaling methods for kinetic potentials, and applications to solid Al and Si," Chem. Phys. Lett. **473**(4), 263–267 (2009).

[25] J.-D. Chai, and J.D. Weeks, "Orbital-free density functional theory: Kinetic potentials and ab initio local pseudopotentials," Phys. Rev. B **75**(20), 205122 (2007).

[26] W.C. Topp, and J.J. Hopfield, "Chemically Motivated Pseudopotential for Sodium," Phys. Rev. B **7**(4), 1295–1303 (1973).

[27] B. Zhou, Y. Alexander Wang, and E.A. Carter, "Transferable local pseudopotentials derived via inversion of the Kohn-Sham equations in a bulk environment," Phys. Rev. B **69**(12), 125109 (2004).

[28] B.G. del Rio, J.M. Dieterich, and E.A. Carter, "Globally-Optimized Local Pseudopotentials for (Orbital-Free) Density Functional Theory Simulations of Liquids and Solids," J. Chem. Theory Comput. **13**(8), 3684–3695 (2017).

[29] Th. Starkloff, and J.D. Joannopoulos, "Local pseudopotential theory for transition metals," Phys. Rev. B **16**(12), 5212–5215 (1977).

[30] Q. Xu, C. Ma, W. Mi, Y. Wang, and Y. Ma, "Nonlocal pseudopotential energy density functional for orbital-free density functional theory," Nat. Commun. **13**(1), 1385 (2022).

[31] Y.-C. Chi, and C. Huang, "High-Quality Local Pseudopotentials for Metals," J. Chem. Theory Comput. **20**(8), 3231–3241 (2024).
Page 27 of 30

# Supporting Information

# A machine-learned kinetic energy model for light weight metals and compounds of group III-V elements

Johann Lüder[1,2,3], Manabu Ihara[4], Sergei Manzhos[4,1]

[1] Department of Materials and Optoelectronic Science, National Sun Yat-sen University, 80424, No. 70, Lien-Hai Road, Kaohsiung, Taiwan

[2] Center of Crystal Research, National Sun Yat-sen University, 80424, No. 70, Lien-Hai Road, Kaohsiung, Taiwan

[3] Center for Theoretical and Computational Physics, National Sun Yat-Sen University, Kaohsiung 80424, Taiwan

[4] School of Materials and Chemical Technology, Tokyo Institute of Technology, Ookayama 2-12-1, Meguro-ku, Tokyo 152-8552 Japan

## Contents

Figure S1. Energy-volume curves for binary compounds not shown in the main text. The abscissa is relative strain $V/V_0$ (where $V_0$ is equilibrium volume) and the ordinate axis is energy in $a.u.$. The blue dots are reference DFT data and red (training set) and green (test set) dots are values predicted by the GPR model. Atom labels and Materialsproject ID numbers are shown on the graphs in black. Numbers in blue are the values of $B'$ from reference DFT calculation, and those in red are the values of $B'$ predicted by the ML model.

[1] Author to whom correspondence should be addressed. Email: Manzhos.s.aa@m.titech.ac.jp

Figure S2. Energy-volume curves for unary compounds. The abscissa is relative strain $V/V_0$ (where $V_0$ is equilibrium volume) and the ordinate axis is energy in *a.u*. The blue dots are reference DFT data and red (training set) and green (test set) dots are values predicted by the polynomial model with $p = 4$. Atom labels and Materialsproject ID numbers are shown on the graphs in black. Numbers in blue are the values of $B'$ from reference DFT calculation, and those in red are the values of $B'$ predicted by the ML model.

Figure S3. Energy-volume curves for binary compounds. The abscissa is relative strain $V/V_0$ (where $V_0$ is equilibrium volume) and the ordinate axis is energy in *a.u*. The blue dots are reference DFT data and red (training set) and green (test set) dots are values predicted by the polynomial model with $p = 4$. Atom labels and Materialsproject ID numbers are shown on the graphs in black. Numbers in blue are the values of $B'$ from reference DFT calculation, and those in red are the values of $B'$ predicted by the ML model.

Figure S4. Energy-volume curves for ternary compounds. The abscissa is relative strain $V/V_0$ (where $V_0$ is equilibrium volume) and the ordinate axis is energy in *a.u*. The blue dots are reference DFT data and red (training set) and green (test set) dots are values predicted by the polynomial model with $p = 4$. Atom labels and Materialsproject ID numbers are shown on the graphs in black. Numbers in blue are the values of $B'$ from reference DFT calculation, and those in red are the values of $B'$ predicted by the ML model.

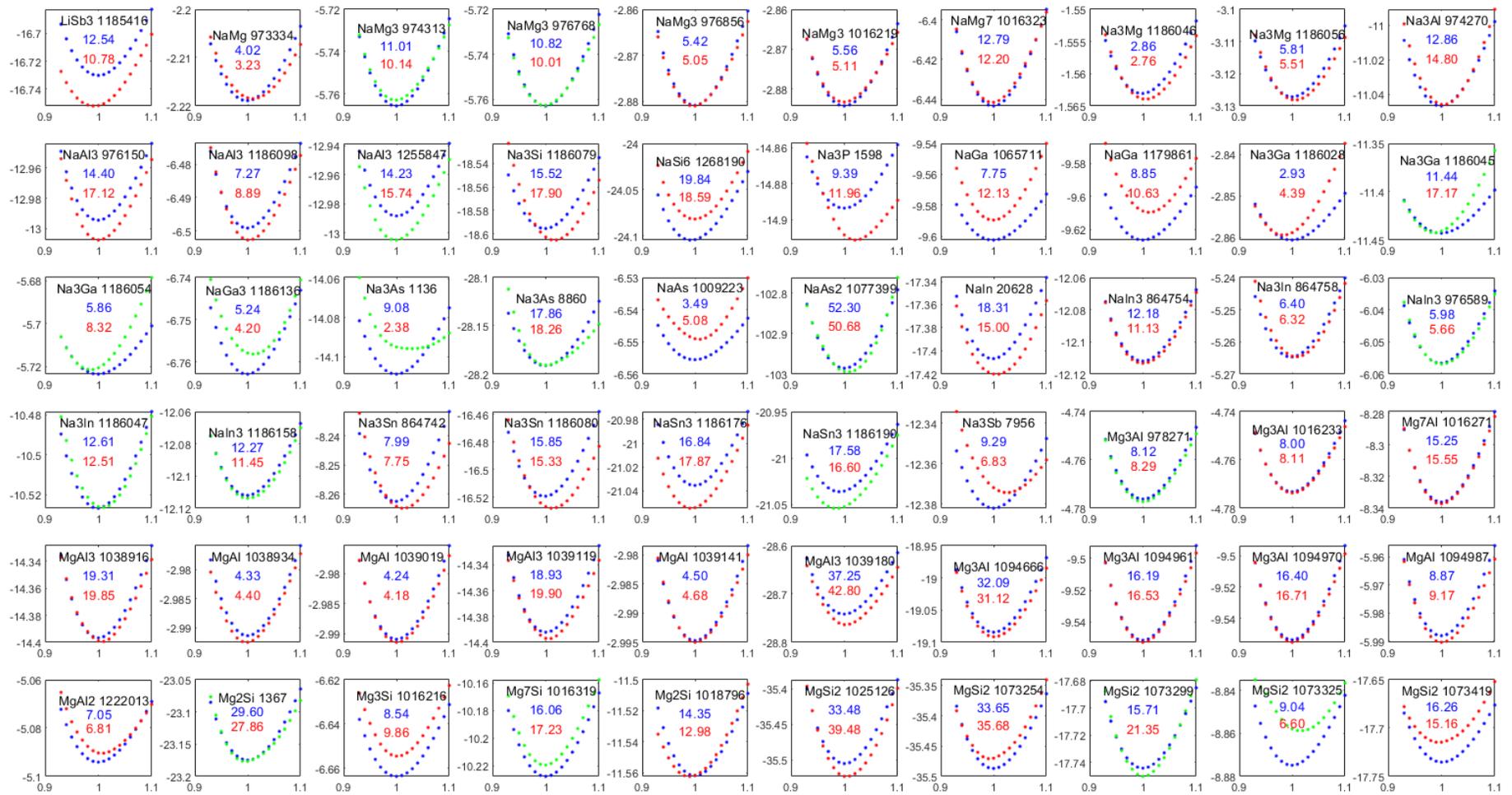

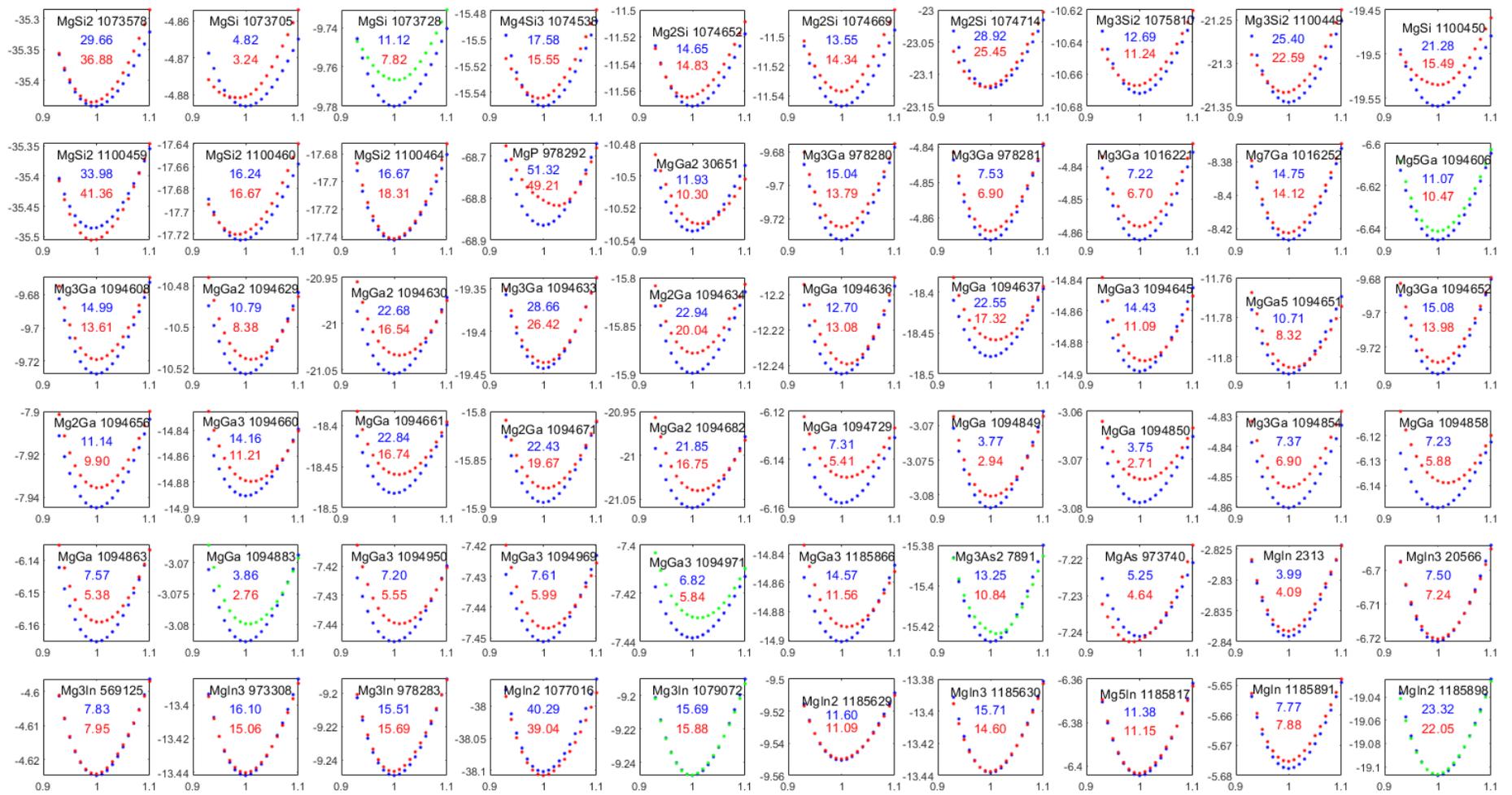

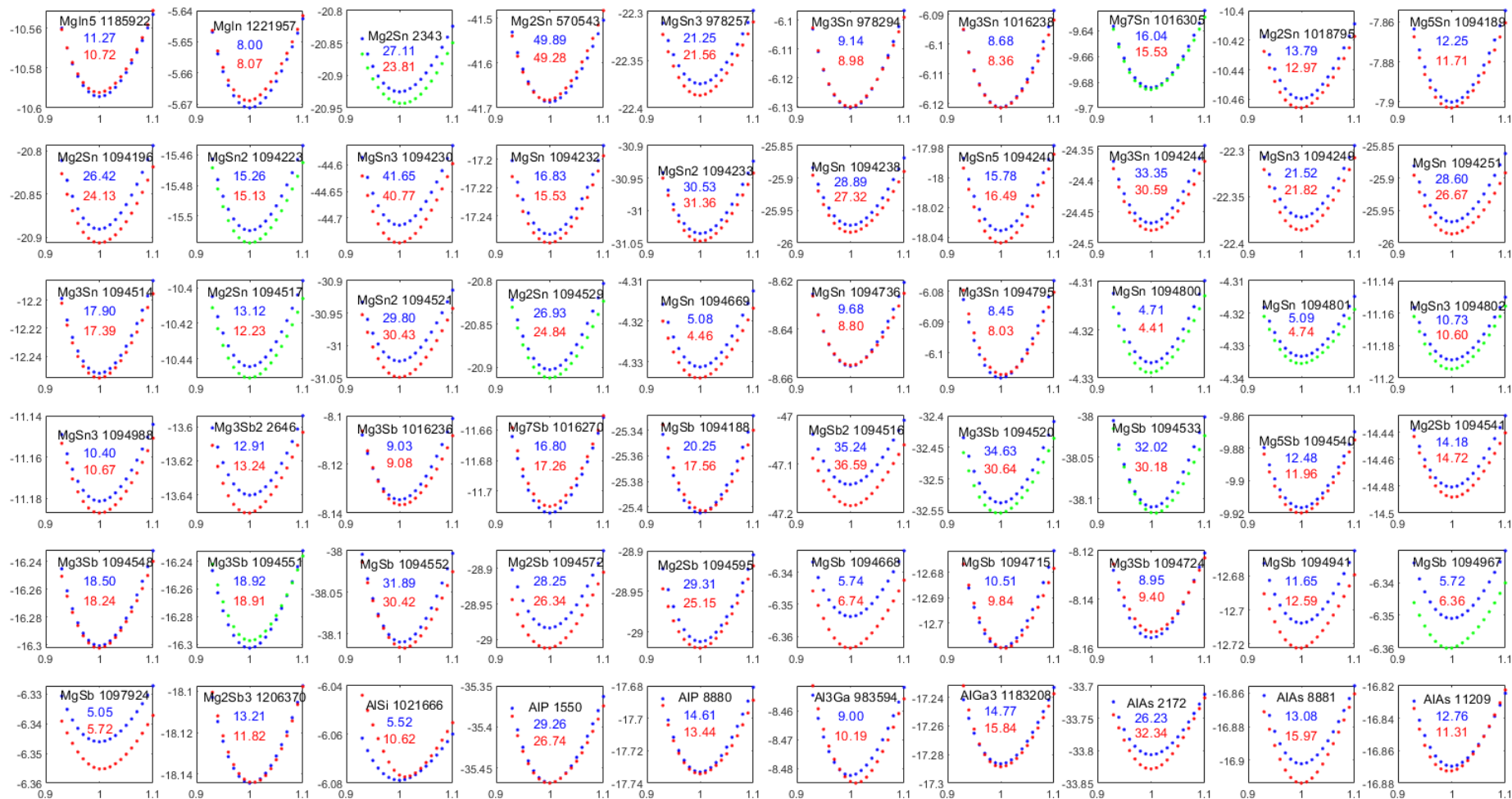

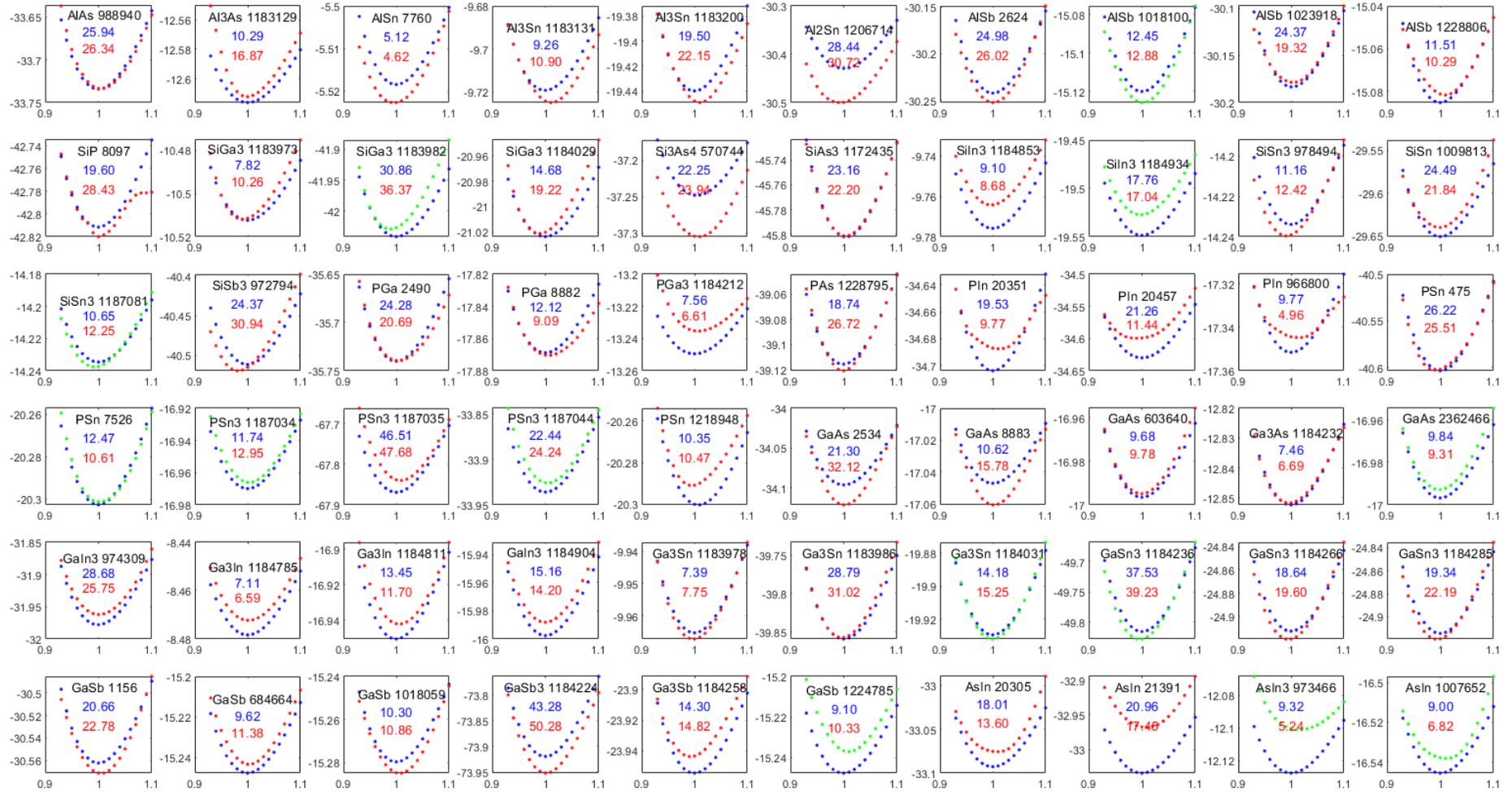

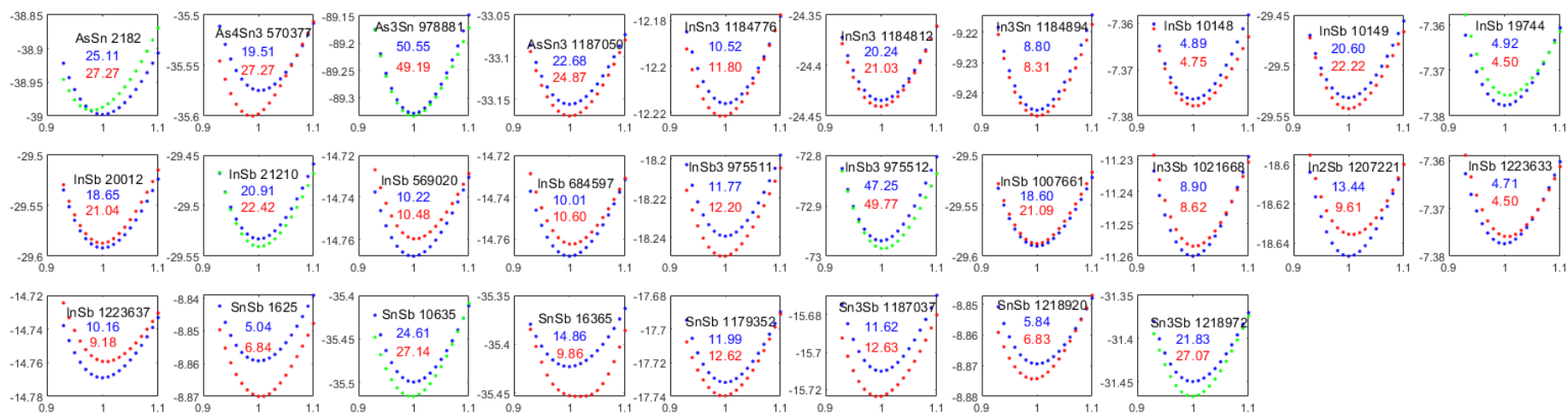

Figure S1. Energy-volume curves for binary compounds not shown in the main text. The abscissa is relative strain $V/V_0$ (where $V_0$ is equilibrium volume) and the ordinate axis is energy in $a.u$. The blue dots are reference DFT data and red (training set) and green (test set) dots are values predicted by the GPR model. Atom labels and Materialsproject ID numbers are shown on the graphs in black. Numbers in blue are the values of $B'$ from reference DFT calculation, and those in red are the values of $B'$ predicted by the ML model.

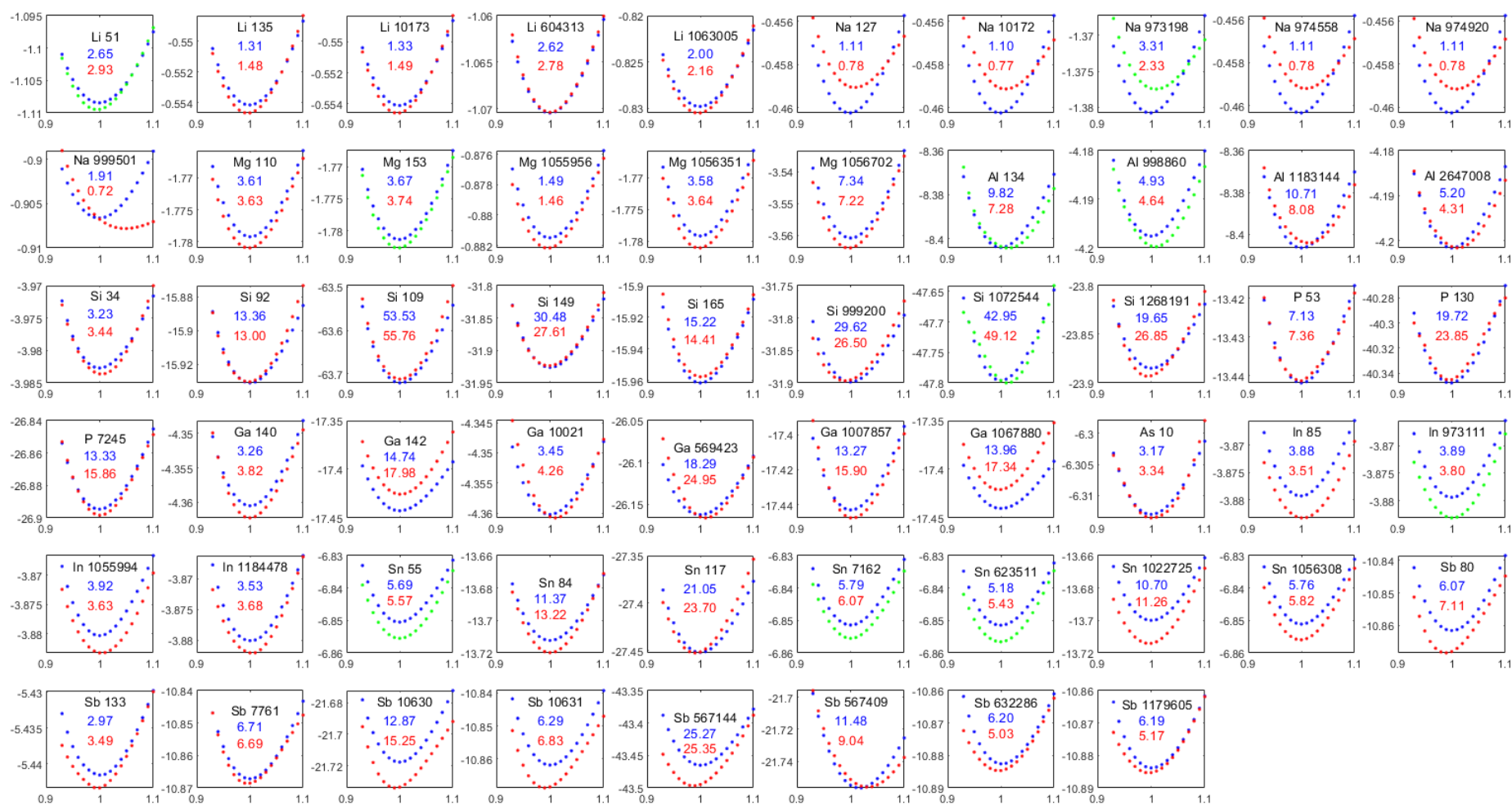

Figure S2. Energy-volume curves for unary compounds. The abscissa is relative strain $V/V_0$ (where $V_0$ is equilibrium volume) and the ordinate axis is energy in *a.u*. The blue dots are reference DFT data and red (training set) and green (test set) dots are values predicted by the polynomial model with $p = 4$. Atom labels and Materialsproject ID numbers are shown on the graphs in black. Numbers in blue are the values of $B'$ from reference DFT calculation, and those in red are the values of $B'$ predicted by the ML model.

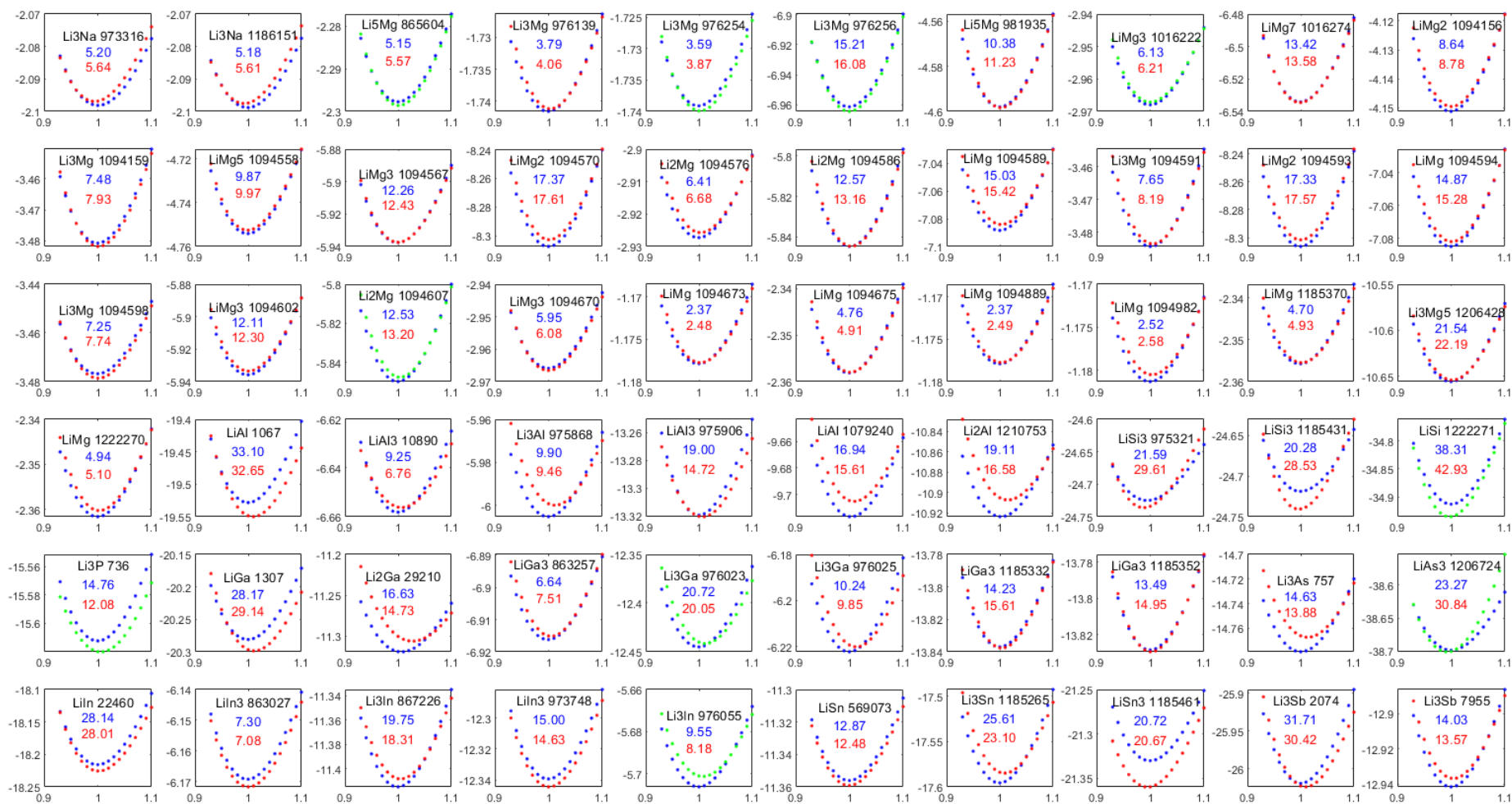

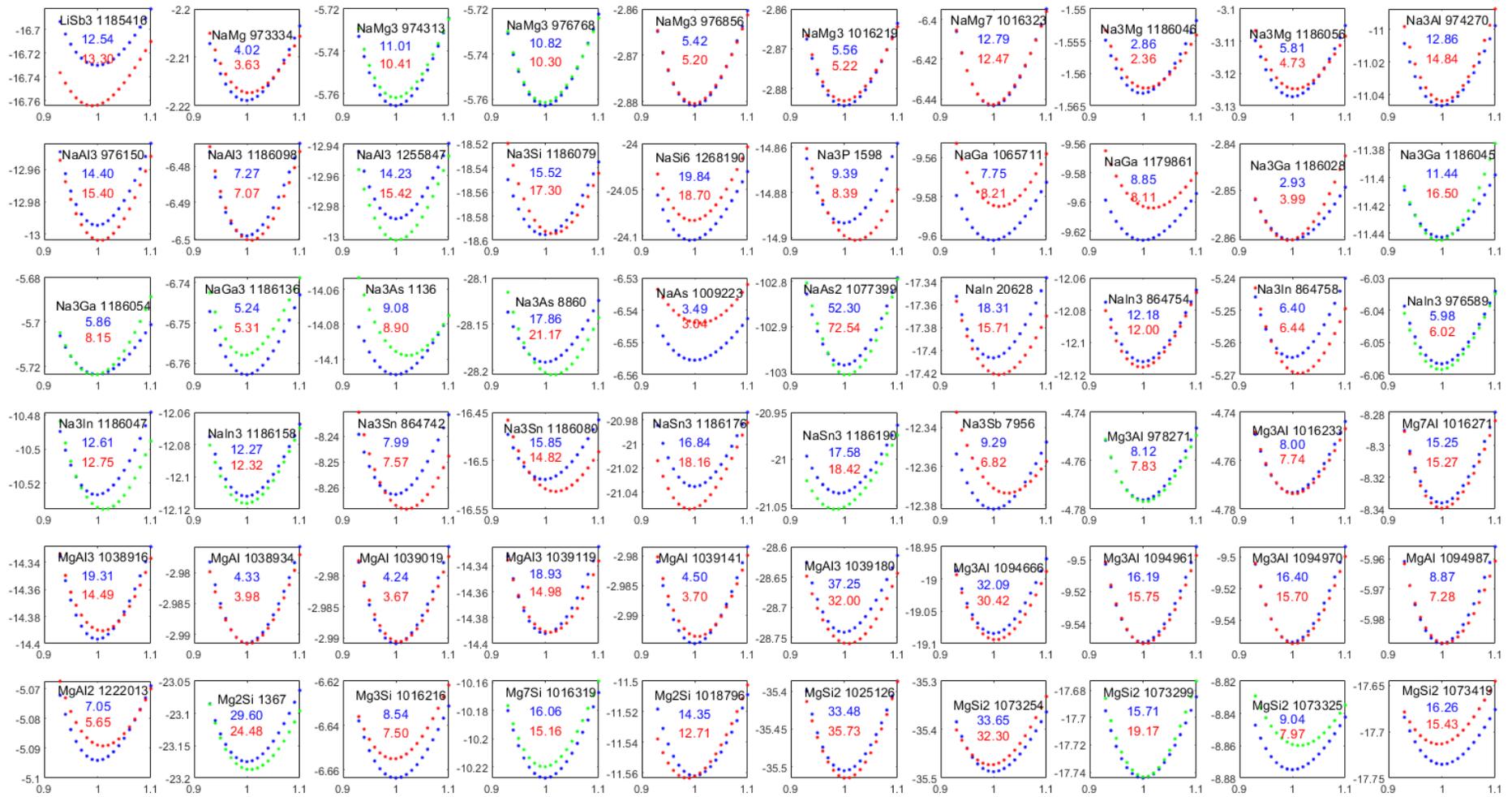

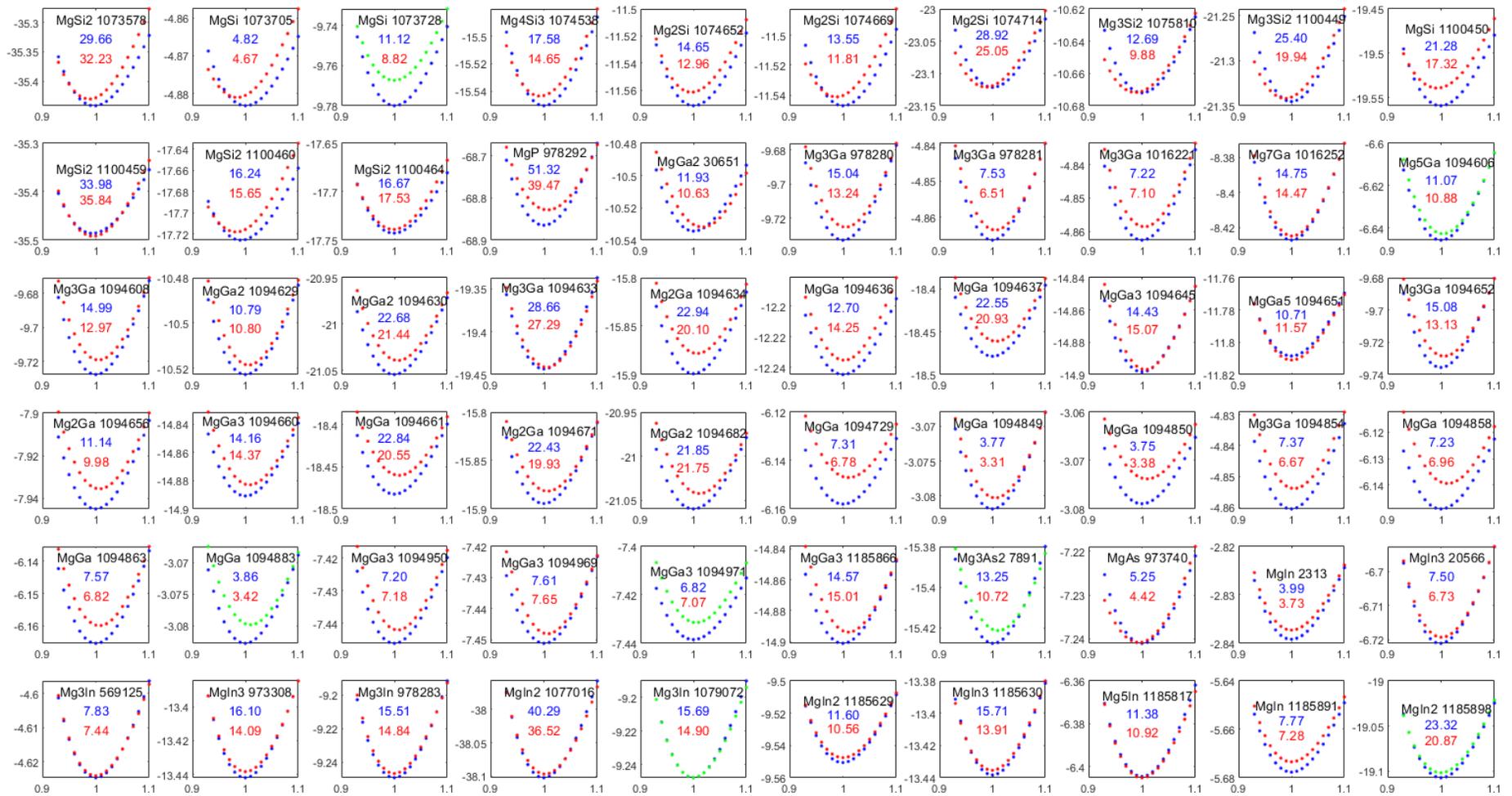

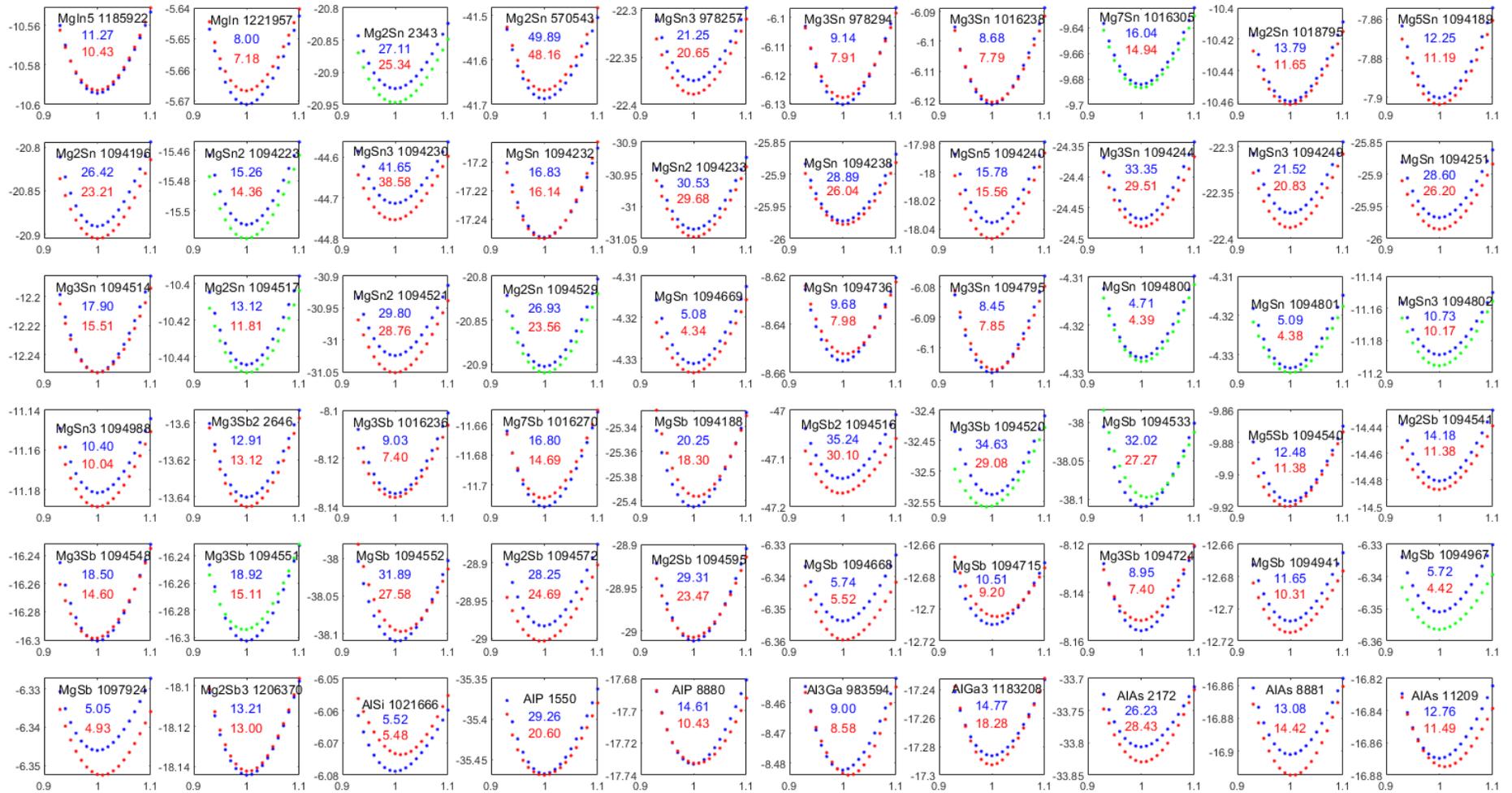

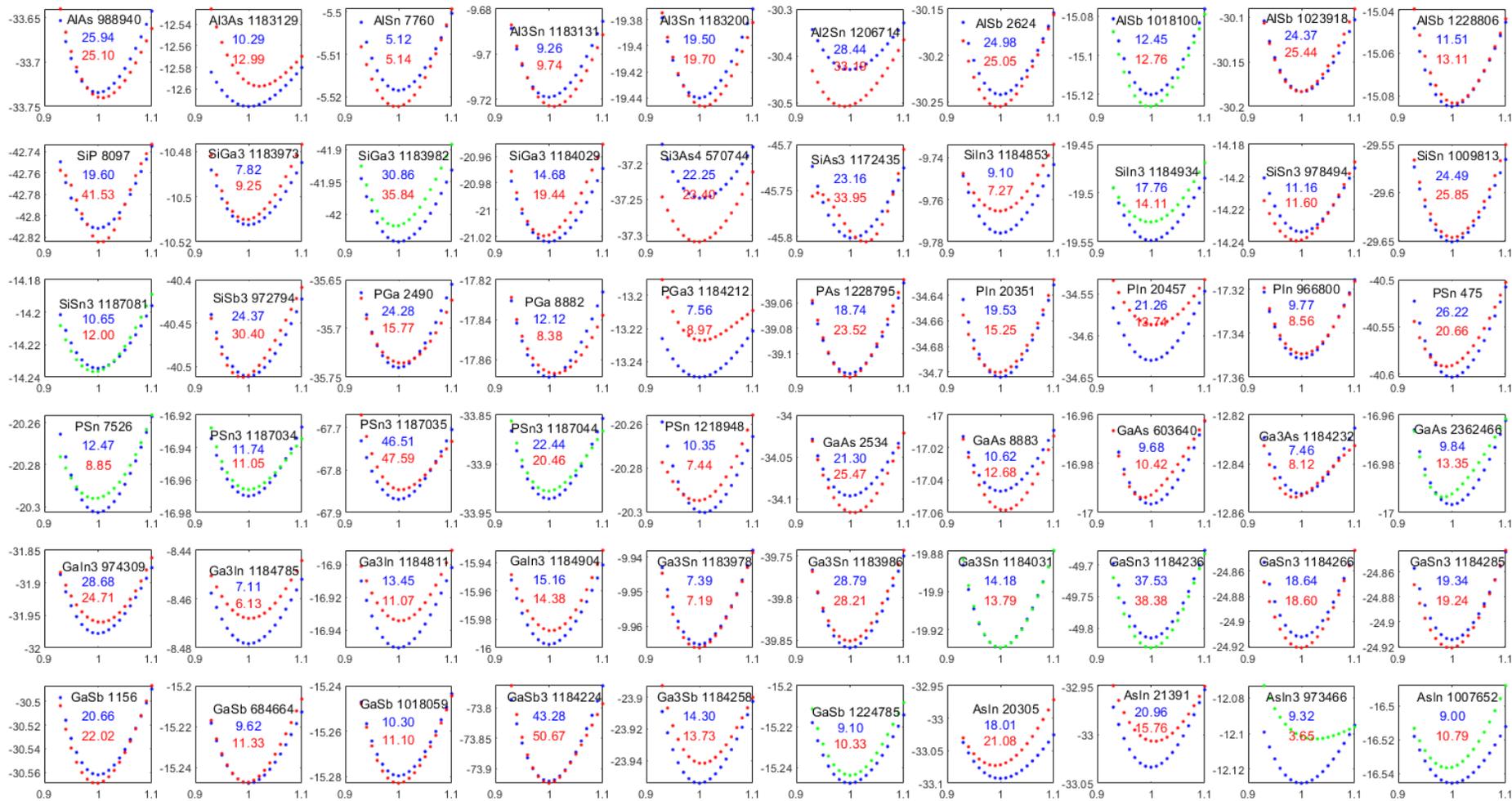

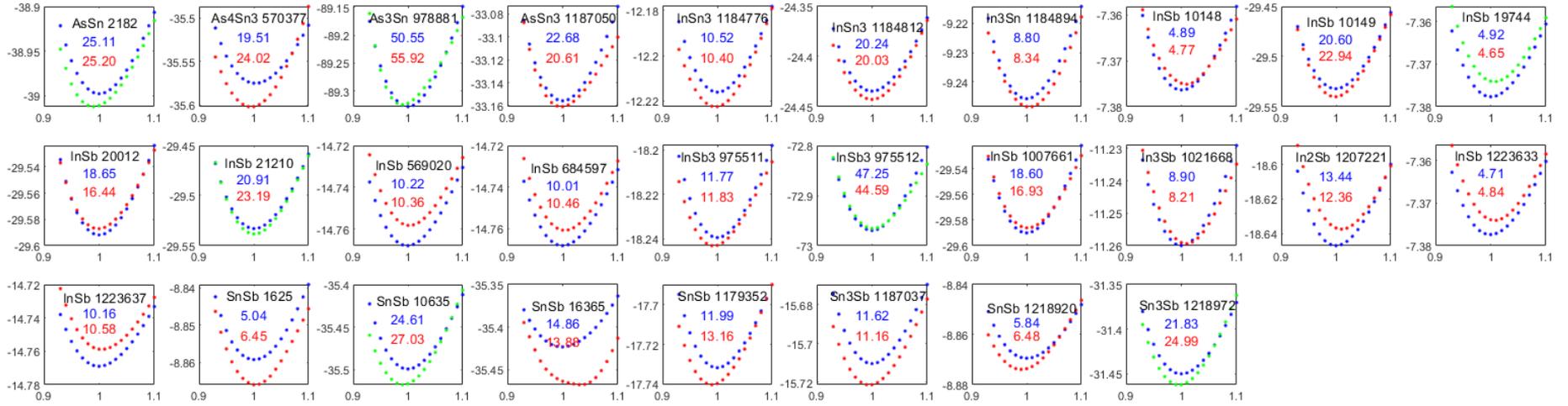

Figure S3. Energy-volume curves for binary compounds. The abscissa is relative strain $V/V_0$ (where $V_0$ is equilibrium volume) and the ordinate axis is energy in *a.u*. The blue dots are reference DFT data and red (training set) and green (test set) dots are values predicted by the polynomial model with $p = 4$. Atom labels and Materialsproject ID numbers are shown on the graphs in black. Numbers in blue are the values of $B'$ from reference DFT calculation, and those in red are the values of $B'$ predicted by the ML model.

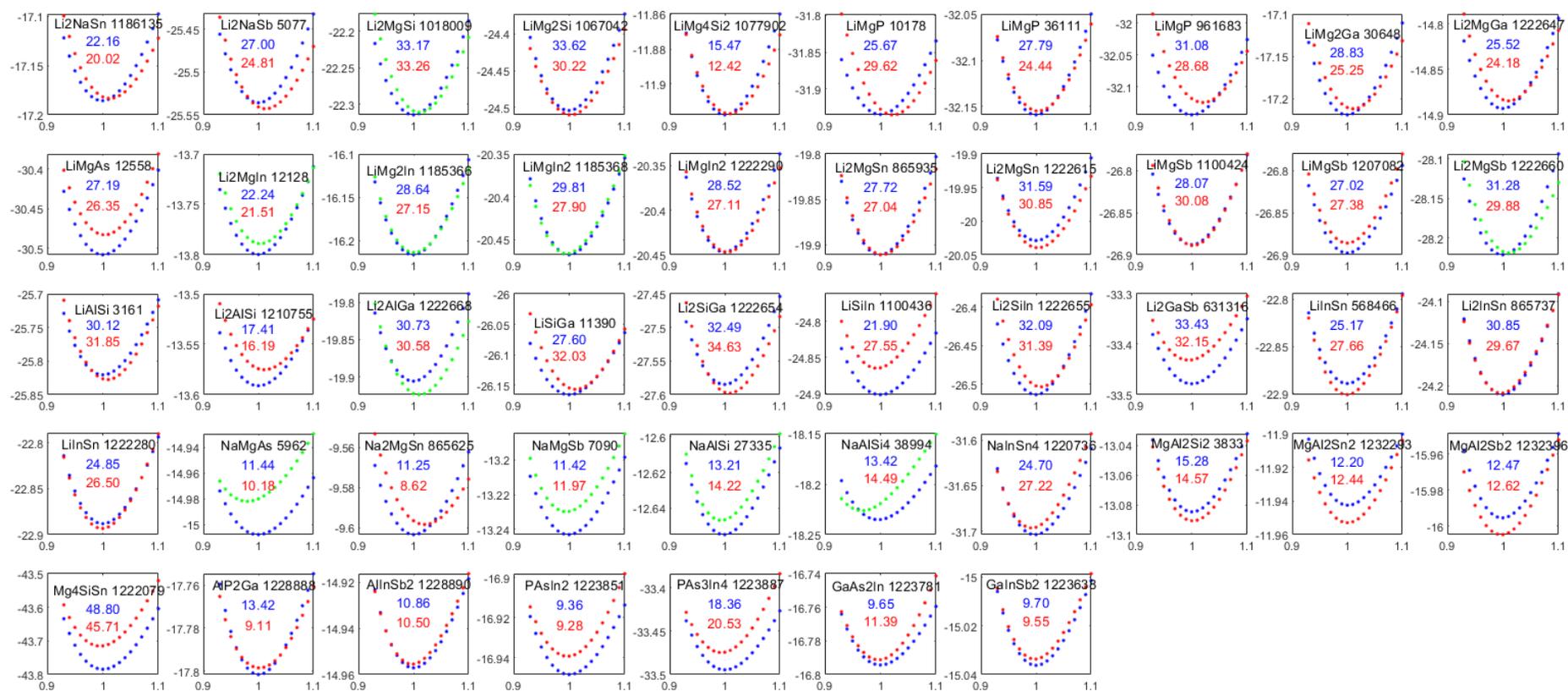

Figure S4. Energy-volume curves for ternary compounds. The abscissa is relative strain $V/V_0$ (where $V_0$ is equilibrium volume) and the ordinate axis is energy in *a.u.* The blue dots are reference DFT data and red (training set) and green (test set) dots are values predicted by the polynomial model with $p = 4$. Atom labels and Materialsproject ID numbers are shown on the graphs in black. Numbers in blue are the values of $B'$ from reference DFT calculation, and those in red are the values of $B'$ predicted by the ML model.